\documentclass[twocolumn]{aastex61}

\newcommand\aastex{AAS\TeX}

\received{\today}
\revised{\***, 2017}
\accepted{\***,2017}
\submitjournal{ApJ}

\shorttitle{\aastex\ ALMA observations of G9.62+0.19}
\shortauthors{Liu et al.}

\shorttitle{ALMA reveals sequential high-mass star formation in the G9.62+0.19 complex } \shortauthors{Liu et al.}

\begin{document}

\title{ALMA reveals sequential high-mass star formation in the G9.62+0.19 complex}
\correspondingauthor{Tie Liu}
\email{liutiepku@gmail.com}

\author{Tie Liu}
\affiliation{Korea Astronomy and Space Science Institute 776, Daedeokdae-ro, Yuseong-gu, Daejeon, Republic of Korea (34055)}
\affiliation{Chinese Academy of Sciences, South America Center for Astrophysics (CASSACA) at Cerro Cal\'{a}n, Camino El Observatorio \#1515, Las Condes, Santiago, Chile}
\affiliation{Department of Astronomy, Peking University, 100871, Beijing, China}
\affiliation{EAST ASIAN OBSERVATORY, 660 N. A'oh\={o}k\={u} Place, Hilo, Hawaii 96720-2700, USA}

\author{John Lacy}
\affiliation{Department of Astronomy, University of Texas at Austin, Austin, TX 78712, USA}

\author{Pak Shing Li}
\affiliation{Astronomy Department, University of California, Berkeley, CA 94720}

\author{Ke Wang}
\affiliation{European Southern Observatory, Karl-Schwarzschild-Str.2, D-85748 Garching bei M\"{u}nchen, Germany}

\author{Sheng-Li Qin}
\affiliation{Department of Astronomy, Yunnan University, and Key Laboratory of Astroparticle Physics of Yunnan Province, Kunming, 650091, China}

\author{Qizhou Zhang}
\affiliation{Harvard-Smithsonian Center for Astrophysics, 60 Garden Street, Cambridge, MA 02138, USA}

\author{Kee-Tae Kim}
\affiliation{Korea Astronomy and Space Science Institute 776, Daedeokdae-ro, Yuseong-gu, Daejeon, Republic of Korea (34055)}

\author{Guido Garay}
\affiliation{Departamento de Astronom\'{\i}a, Universidad de Chile, Casilla 36-D, Santiago, Chile}

\author{Yuefang Wu}
\affiliation{Department of Astronomy, Peking University, Beijing 100871, China}

\author{Diego Mardones}
\affiliation{Departamento de Astronom\'{\i}a, Universidad de Chile, Casilla 36-D, Santiago, Chile}

\author{Qingfeng Zhu}
\affiliation{Astronomy Department, University of Science and Technology, Chinese Academy of Sciences, Hefei 210008, China}

\author{Ken'ichi Tatematsu}
\affiliation{National Astronomical Observatory of Japan, 2-21-1 Osawa, Mitaka, Tokyo 181-8588, Japan}

\author{Tomoya Hirota}
\affiliation{National Astronomical Observatory of Japan, 2-21-1 Osawa, Mitaka, Tokyo 181-8588, Japan}

\author{Zhiyuan Ren}
\affiliation{National Astronomical Observatories, Chinese Academy of Science, A20 Datun Road, Chaoyang District, Beijing 100012, China}

\author{Sheng-Yuan Liu}
\affiliation{Academia Sinica, Institute of Astronomy and Astrophysics, P.O. Box 23-141, Taipei 106, Taiwan}

\author{Huei-Ru Chen}
\affiliation{Academia Sinica, Institute of Astronomy and Astrophysics, P.O. Box 23-141, Taipei 106, Taiwan}

\author{Yu-Nung Su}
\affiliation{Academia Sinica, Institute of Astronomy and Astrophysics, P.O. Box 23-141, Taipei 106, Taiwan}

\author{Di Li}
\affiliation{National Astronomical Observatories, Chinese Academy of Science, A20 Datun Road, Chaoyang District, Beijing 100012, China}
\affiliation{Key Laboratory for Radio Astronomy, Chinese Academy of Sciences, Nanjing 210008, China}

\begin{abstract}

Stellar feedback from high-mass stars (e.g., H{\sc ii} regions) can strongly influence the surrounding interstellar medium and regulate star formation. Our new ALMA observations reveal sequential high-mass star formation taking place within one sub-virial filamentary clump (the G9.62 clump) in the G9.62+0.19 complex. The 12 dense cores (MM 1-12) detected by ALMA are at very different evolutionary stages, from starless core phase to UC H{\sc ii} region phase. Three dense cores (MM6, MM7/G, MM8/F) are associated with outflows. The mass-velocity diagrams of outflows associated with MM7/G and MM8/F can be well fitted with broken power laws. The mass-velocity diagram of SiO outflow associated with MM8/F breaks much earlier than other outflow tracers (e.g., CO, SO, CS, HCN), suggesting that SiO traces newly shocked gas, while the other molecular lines (e.g., CO, SO, CS, HCN) mainly trace the ambient gas continuously entrained by outflow jets. Five cores (MM1, MM3, MM5, MM9, MM10) are massive starless core candidates whose masses are estimated to be larger than 25 M$_{\sun}$, assuming a dust temperature of $\leq$ 20 K. The shocks from the expanding H{\sc ii} regions (``B" \& ``C") to the west may have great impact on the G9.62 clump through compressing it into a filament and inducing core collapse successively, leading to sequential star formation. Our findings suggest that stellar feedback from H{\sc ii} regions may enhance the star formation efficiency and suppress the low-mass star formation in adjacent pre-existing massive clumps.

\end{abstract}


\section{Introduction}

As the principal sources of heavy elements and UV radiation
\citep{zin07}, high-mass stars play a major role in the evolution of galaxies.
However, the formation and evolution of high-mass stars are still unclear. There are two most promising models to account for high-mass star formation. One is called Turbulent Core Accretion \citep{york02,mck03,kru05,kru06}, and the other Competitive Accretion \citep{bon96,bonn02,bonn06,bonn08}. The difference between the two models lies primarily in how and when the mass is gathered to form the massive star. The former suggests that high-mass stars form directly from isolated massive gas cores, as do isolated low-mass stars, but with a much larger accretion rate. The latter claims that the initial fragmentation of a clump results in dense cores that have masses typical of the thermal Jeans, while high-mass stars form in the central cores through competing for interclump gas with the other off-center cores. In the Turbulent Core Accretion model, the individual cores are the gas reservoir and will gravitationally collapse, but the bulk of the gas will not. Therefore, the Turbulent Core Accretion model predicts the existence of massive starless cores to form high-mass stars through monolithic collapse. In contrast, the Competitive Accretion model predicts the cloud gas is free to be accreted due to a common potential. Thus global collapse at clump/cloud scale may happen in the Competitive Accretion model but not in the Turbulent Core Accretion model. The discovery of global collapse in highly fragmented massive clumps with the most massive cores residing at their center may support the Competitive Accretion model \citep[e.g.,][]{liu13a,liu13b,pere13,zhang09,zhang11,zhang15}. In contrast, the detection of Keplerian-like disks around the forming OB stars strongly indicates that high-mass stars may form in a similar way as their low-mass counterparts \citep[e.g.,][]{zhang98,keto10,john15,chen16}.

Although very promising, neither the Turbulent Core Accretion model nor the Competitive Accretion model takes account of stellar feedback from massive stars, which can strongly influence the surrounding interstellar medium and regulate star formation through photoionizing radiation, energetic winds, or supernova explosions. The expansion of shock front (SF) that emerges around massive stars can compress the ISM, triggering star formation in very dense layers \citep{el77,whit94a,whit94b}. This so-called ``collect and collapse" process can self propagate and lead to sequential star formation \citep{el77,whit94a,whit94b}. The ``collect and
collapse" process has been revealed in the borders of several H{\sc ii} regions evidenced by fragmented shells or sequential star formation \citep{de03,de05,za06,za07,de08,po09,pe10,bra11,liu12,liu15,liu16a}. \cite{thom12} estimated that the fraction of massive stars in the Milky Way formed by triggering processes could be between 14 and 30 per cent through studying a large sample of infrared bubbles. However, recent numerical simulations indicate that the ages and geometrical distribution of stars relative to the feedback source or feedback-driven structure (e.g. shells, pillar structures), may be not substantially helpful in distinguishing triggered star formation from spontaneous star formation \citep{dale15}. In addition, in most of those previous studies, the fragmented shells surrounding the H{\sc ii} regions are not dense enough to form a new generation of high-mass stars. Can high-mass star formation be triggered? To answer this question, detailed studies of massive clumps (the birth places of high-mass stars) near feedback sources (H{\sc ii} regions or Supernova remnants) are urgently needed.

\subsection*{The G9.62+0.19 complex}

Located at a distance of 5.2 kpc \citep{san09}, the G9.62+0.19 complex contains a cluster of radio
continuum sources (denoted from A-I), which are at different evolutionary stages \citep[e.g., Extended H{\sc ii} region ``A", Cometary-shaped H{\sc ii} region ``B", compact H{\sc ii} region ``C", Ultra compact H{\sc ii} region ``D \& E" and hot molecular core ``F";][]{gar93,tes00,liu11}.
Sequential high-mass star formation is taking place in the G9.62+0.19 complex \citep{hof94,hof96,hof01,tes00,liu11}. The youngest sources are located in a clump to the east of the evolved H{\sc ii} regions \citep{liu11}. Therefore, the G9.62+0.19 complex is an ideal target to study the effect of stellar feedback on new generations of high-mass star formation.

\begin{figure}[tbh!]
\centering
\includegraphics[angle=0,scale=0.4]{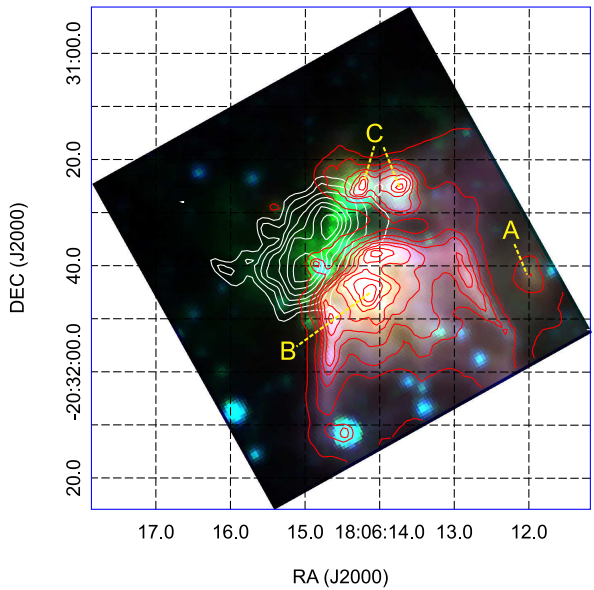}
\caption{Spitzer/IRAC 3-color composite image (8 $\micron$ in red, 4.5 $\micron$ in green, and 3.6 $\micron$ in blue) of the G9.62+0.19 complex. The Spitzer/IRAC 8 $\micron$ emission is also shown in red contours. The contour levels are [0.1, 0.2, 0.3,0.4,0.5, 0.6, 0.7,0.8,0.9]$\times$2550 MJy/sr.
The JCMT/SCUBA 450 $\micron$ emission from the G9.62 clump is shown in white contours. The contour levels are [0.3, 0.4, 0.5, 0.6, 0.7, 0.8, 0.9]$\times$96 Jy/beam. A, B and C are the three evolved H{\sc ii} regions. \label{ds9}}
\end{figure}

Figure \ref{ds9} shows the Spitzer/IRAC three color (8 $\micron$ in red, 4.5 $\micron$ in green, and 3.6 $\micron$ in blue) composite image of the G9.62+0.19 complex. The red contours represent the Spitzer/IRAC 8 $\micron$ polycyclic aromatic hydrocarbon (PAH) emission, which traces the photodissociation regions (PDRs) of the evolved H{\sc ii} regions \citep[radio sources ``A", ``B" and ``C",][]{gar93}. The white contours show the 450 $\micron$ continuum emission from JCMT/SCUBA, which traces the dust emission from the new high-mass star forming region \citep[e.g.,][]{hof94,hof96,hof01,tes00,liu11}. Hereafter we call the region traced by 450 $\micron$ continuum emission the G9.62 clump. New generations of high-mass young stellar objects (YSOs) are forming in the G9.62 clump. The G9.62 clump has a mean number density of $\sim(9.1\pm0.7)\times10^{4}$ cm$^{-3}$, a mass of $\sim2800\pm200$ M$_{\sun}$ and a luminosity of $\sim(1.7\pm0.1)\times10^6$ L$_{\sun}$ (see APPENDIX A). The G9.62 clump is located to the east of the evolved H{\sc ii} regions and is surrounded by their PDRs. In addition, the G9.62 clump is associated with extended 4.5 $\micron$ emission, which may indicate the existence of shocked H$_{2}$ emission. The G9.62 clump may be compressed by the shocks induced from the evolved H{\sc ii} region. It is very clear that large-scale (parsec scale) sequential high-mass star formation is taking place from west to east in the G9.62+0.19 complex \citep[e.g.,][]{liu11}.

In this work, we study the interaction between the evolved H{\sc ii} regions (``A", ``B" \& ``C") and the G9.62 clump. Particularly, we thoroughly investigate the fragmentation and outflows in the G9.62 clump from the high resolution and high sensitivity ALMA observations.

\section{Observations}

\subsection{ALMA Observations}

The observations (Project ID: 2013.1.00957.S) of the G9.62+0.19 complex were conducted with ALMA on 2015 April 27 in its compact configuration with a total of 39 antennas in the 12-m array; on 2015 May 24 in its extended configuration with a total of 34 antennas in the 12-m array; and on 2015 May 04 with 10 antennas in the 7-m array. Quasar J1733-1304 was observed for phase and bandpass calibrations in the observations with the extended configuration of 12-m array and also in the observations with the 7-m array. In the observations with the compact configuration of the 12-m array, Quasars J1517-2422 and J1733-1304 were used for bandpass calibration and phase calibration, respectively. Neptune, Titan and Quasar J1733-130 were used for flux calibration in the observations with the extended configuration of the 12-m array, compact configuration of the 12-m array and the 7-m array, respectively. We used two pointings to cover the whole region of the target (see Figure \ref{obs} in APPENDIX). One phase reference center was R.A.(J2000)=$18^h06^m14^s.67$ and decl.(J2000)=$-20\arcdeg31\arcmin31\arcsec.91$, and the other was R.A.(J2000)=$18^h06^m14^s.90$ and  decl.(J2000)=$-20\arcdeg31\arcmin40\arcsec.21$. The observations employed the Band 6 (230 GHz) receivers with dual-polarization mode. We used five spectral windows with a bandwidth of 117 MHz in each window and another spectral window with a bandwidth of 234 MHz to cover multiple lines such as CO (2-1), SiO (5-4), CH$_{3}$OH v$_{t}$=1 ($6_{1,5}-7_{2,6}$) and H${30\alpha}$. The spectral windows have a uniform channel width of 61 KHz (or 0.08 km~s$^{-1}$), which is chosen to resolve any velocity gradients caused by bulk motions or keplerian rotation. We smoothed the spectra lines to 0.16 km~s$^{-1}$ spectral resolution in data analysis, which is high enough to resolve line profiles with expected linewidths of $\sim$5 km~s$^{-1}$ \citep{liu11}.

The visibility data were calibrated in CASA by the ALMA supporting staff. We examined the calibration. We used updated antenna positions and fixed resolved flux calibrators. The visibility data from the three configurations were combined in CASA. The largest recoverable scale (LAS) in the combined data is $\sim20\arcsec$, which is much larger than the typical core sizes ($\sim$0.1 pc; or 4$\arcsec$ at 5.2 kpc). The LAS is also larger than the outflow extends ($\sim10\arcsec$, see section 3.3) in the G9.62 clump. Therefore, although no single dish data included in the combination, the missing flux is not a severe problem for analyzing the properties of dense cores or outflows in this work. We construct continuum visibility data using the line-free spectral channels with a total bandwidth of 350 MHz and applied self-calibration on the continuum in order to further improve the dynamical range in the maps. The gain solutions from the self-calibration were applied to the spectral line data. The continuum image reaches a 1$\sigma$ rms noise of 0.8 mJy in a synthesized beam of $0\arcsec.94\times0\arcsec.71$ (P.A.=80.25 deg). For molecular lines, we only focused on SiO (5-4), CO (2-1), and CH$_{3}$OH v$_{t}$=1 ($6_{1,5}-7_{2,6}$) in this work. The rest frequencies of SiO (5-4), CO (2-1), and CH$_{3}$OH v$_{t}$=1 ($6_{1,5}-7_{2,6}$) lines are 217.10498, 230.53800 and 217.29920 GHz, respectively \citep[obtained from Splatalogue catalog; for line identification, we refer to][]{zer12}. The synthesized beams for SiO (5-4), CO (2-1) and CH$_{3}$OH v$_{t}$=1 ($6_{1,5}-7_{2,6}$) lines are $1\arcsec.02\times0\arcsec.75$ (P.A.=84.97 deg), $0\arcsec.94\times0\arcsec.73$ (P.A.=87.02 deg) and $0\arcsec.99\times0\arcsec.72$ (P.A.=75.42 deg), respectively. The 1$\sigma$ rms noise for lines is $\sim$10 mJy/beam per 0.16 km~s$^{-1}$ channel.

\subsection{[Ne {\sc ii}] observations}

We observed the H{\sc ii} regions in the [Ne {\sc ii}] 12.8 $\micron$ line
with the high spectral resolution, mid-IR(5-25 $\micron$) cross dispersed
spectrograph TEXES \citep[Texas Echelon Cross Echelle Spectrograph,][]{lacy02} on the 3.0 NASA IRTF on Mauna Kea in Hawaii on July 2, 2014. The region covered by [Ne {\sc ii}] slits is shown in Figure \ref{obs}.
The observations were carried out in the high resolution mode of
the instrument with the spectral resolution approximately 4 km~s$^{-1}$. A 1$\arcsec$ wide, 10$\arcsec$ long
slit was oriented along the north-south direction in the observations.
We scanned the object from east to west by stepping the
telescope with the step size of 0.7$\arcsec$. There are 121 steps in each scan.
The scan length
was determined so that the telescope pointing was considered safely far
from the H{\sc ii} region at the both ends of the scans and additional steps at
the beginning and the end of the scans are available and can be used
to measure the sky background level and to do sky subtraction.
Ten north-south partially overlapping scans were made.
They are offset north-south by about 5$\arcsec$.
The exposure time for each step in the scans is 8 seconds.
Due to the un-equal coverage of the declination,
the total exposure time for any position in the field ranges between
8 to 32 seconds. TEXES measures ambient background and sky flats before each
scan for flux calibration. We should point out that the sky background subtraction is hard for regions with a large size and weak emission (e.g., radio source ``A"). A drift in the sky background would cause worse residuals in baselines of lines. The spikes on the sides of the [Ne {\sc ii}] lines in C-e and C-w (see Figure \ref{NeII}) are at the wavelengths of telluric lines, so must have been due to drifts in the sky background.

We use a custom FORTAN reduction program \citep{lacy02} to correct optical
distortions, flat-field, cosmic ray (bad pixel) removal, wavelength calibration, and flux calibration. The wavelength calibration is performed with atmospheric lines, whose wavelengths are obtained from HITRAN \citep{roth98}. The sky background at each slit position is linearly interpolated using the sky frames at the
both ends of the scans and subtracted. We then shift and combine multiple scans by using
a cross-correlation algorithm and form a 2D spatial and 1D spectral datacube.
Finally, the datacube is collapsed along the spectral dimension to form a map
of the [Ne {\sc ii}] line emission. The velocity range of the datacube is from -50 to 50 km~s$^{-1}$, which well covers the [Ne {\sc ii}] line emission.

\subsection{JCMT C$^{18}$O (3-2) observations}

The C$^{18}$O (3-2) datacube was obtained from the JCMT archive and was not published in literatures. The observations of C$^{18}$O (3-2) were conducted in May 2011. The HARP, a 16-pixels focal-plane array receiver, was used in the observations of C$^{18}$O (3-2). The main beam efficiency is 0.72. The
Auto-Correlation Spectral Imaging System backend was used in conjunction with HARP and configured to
use a 250 MHz bandwidth with 4096 frequency channels of width 61.0 kHz (or 0.056 km~s$^{-1}$). The rms level per channel is about 0.8 K in brightness temperature.

\section{Results}

\subsection{ALMA 1.3 mm continuum}

\begin{figure*}[tbh!]
\centering
\includegraphics[angle=-90,scale=0.65]{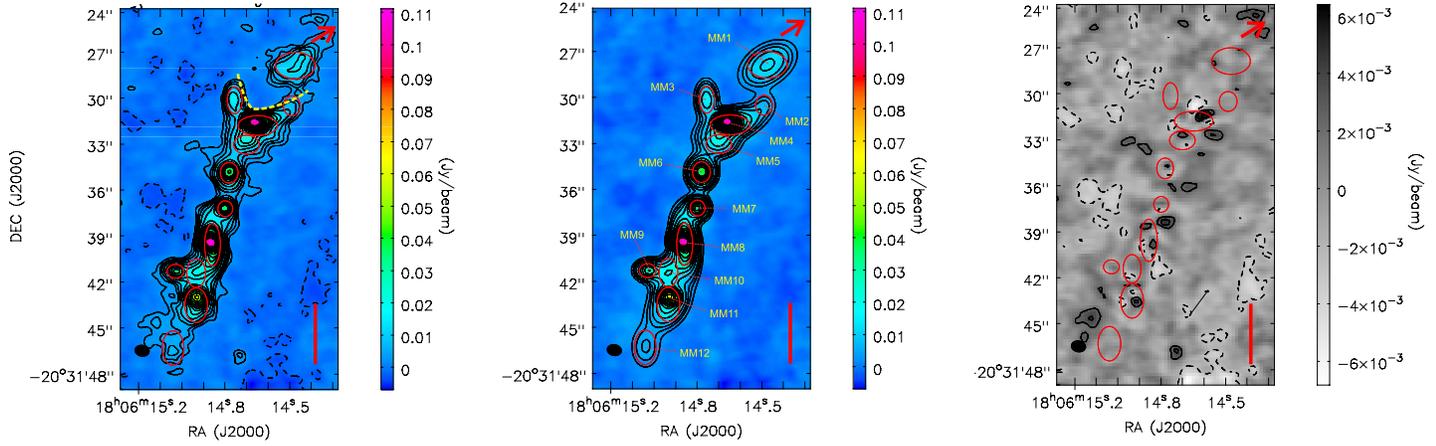}
\caption{The 1.3 mm continuum from ALMA observations. The red ellipses represent the dense cores identified from 2-D Gaussian fit. The black filled ellipse represents the beam. The vertical red line represents a spatial scale of 0.1 pc. The position of radio source ``C" is marked by an arrow. Left panel: observed image, the contours are [-3, 3, 5, 7, 10, 15, 20, 25, 30, 35, 40, 50, 60, 70, 80, 90, 100, 110]$\times$0.8 mJy/beam. As depicted by the yellow dashed curve, the northern part of the G9.62 clump is very likely bent by the compression of the H{\sc ii} region ``C". Middle panel: fitted image, the contours are [3, 5, 7, 10, 15, 20, 25, 30, 35, 40, 50, 60, 70, 80, 90, 100, 110]$\times$0.8 mJy/beam. Right: residual image, the contours are [-3, 3, 5]$\times$0.8 mJy/beam.\label{continuum}}
\end{figure*}

The 1.3 mm continuum of the G9.62 clump from the ALMA observations is shown in the left panel of Figure \ref{continuum}. The G9.62 clump is dominated by a highly fragmented filamentary structure extending from northwest to southeast direction. We identified 12 dense cores above 5 $\sigma$ signal level from the 1.3 mm continuum map. The positions, peak flux, total flux and sizes of these dense cores were obtained from 2-D Gaussian fits and presented in Table \ref{corepara}. However, we found that MM4, MM6, MM8 and MM11 cannot be well fitted with a single gaussian component because point-like source emission remains in the residual images. Therefore, their continuum emission was fitted with a gaussian component plus a point source. The fitted image and residual image from 2-D Gaussian fits and point source fits are presented in the middle panel and right panel of Figure \ref{continuum}, respectively. Only 5 of the 12 dense cores were previously reported. MM4, MM7, MM8 and MM11 are associated with the radio sources ``E", ``G", ``F" and ``D", respectively, which were detected in previous centimeter and millimeter observations \citep[e.g.,][]{tes00,liu11}. MM1 was detected in SMA 860 $\micron$ continuum observations \citep{liu11}. The other dense cores are newly detected in this work. The peak coordinates, peak fluxes (I$_{peak}$), total flux densities (S$_{\nu}$), full width at half maximum (FWHM) deconvolved major sizes (a), minor sizes (b) and position angles (P.A.) of the dense cores are summarized in Table \ref{corepara}. The radii (R) of the dense cores were derived as $R=\sqrt{ab}$ and presented in the 9th column of Table \ref{corepara}. The radii of the dense cores range from 3000 AU to 10000 AU.

The masses of the dense cores can be derived with equations (A1) and (A2) in the appendix. The uncertainties of the masses caused by flux density measurements are negligible as determined from the residual map. The other uncertainties are the dust temperature T$_{d}$ and $\beta$, which are not well known due to the lack of high angular resolution data from other wavelength bands. However, the masses only change $\sim$6\% as $\beta$ changes from 1 to 2, indicating that the uncertainties of masses caused by $\beta$ is also negligible. The main uncertainties of masses come from the determination of T$_{d}$.

We firstly calculate core masses with a T$_{d}$ of 35 K and $\beta$ of 1.5, which are determined from the SED fit for the whole clump (see section 6.2 in the Appendix A). The derived core masses were presented in the 10th column of Table \ref{corepara}. The core masses range from 4 to 87 M$_{\sun}$ with a median value of $\sim$20 M$_{\sun}$. The volume densities of the dense cores range from 0.6$\times10^{6}$ to 2.5$\times10^{7}$ cm$^{-3}$ with a median value of 4.4$\times10^{6}$ cm$^{-3}$. Eleven dense cores are more massive than 10 M$_{\sun}$. There is a lack of a widespread low-mass (M$\leq$ 1-2 M$_{\sun}$) core population in the G9.62 clump. Considering its high mass sensitivity (5 $\sigma$ is $\sim$0.02 M$_{\sun}$ assuming T$_{d}$=35 K and $\beta$=1.5) and high spatial resolution ($\sim$0.025 pc), ALMA should be very easy to detect low mass cores (with typical mass of 1 M$_{\sun}$ and radius of 0.1 pc) if they exist.

However, since MM4, MM7, MM8 and MM11 are either hot cores or Hyper/Ultra compact H{\sc ii} regions with centimeter continuum emission \citep{tes00,liu11} and they also show hot CH$_{3}$OH emission (see next section), their T$_{d}$ should be much higher than the mean value (35 K). Thus, we also calculate their core masses with a T$_{d}$ of $\sim$100 K, which is consistent with the rotational temperature measured from molecular lines of CH$_{3}$CN and H$_{2}$CS \citep{hof96,liu11}. The core masses derived with a T$_{d}$ of $\sim$100 K were shown in parentheses of the 10th column in Table \ref{corepara}. The core masses derived with T$_{d}$ of $\sim$100 K become about one third of the core masses derived with T$_{d}$ of $\sim$35 K. For the other dense cores without radio emission and appear to be at earlier phases (i.e., starless core), we also use a lower dust temperature of 20 K to estimate their core masses, which are shown in parentheses in the 10th column of Table \ref{corepara}. The core masses derived with T$_{d}$ of $\sim$20 K become about two times larger than the core masses derived with T$_{d}$ of $\sim$35 K.

\begin{longrotatetable}
\begin{deluxetable}{ccccccccccccc}
\centering
\scriptsize
\tablecolumns{12} \tablewidth{0pc}
\tablecaption{Parameters of the continuum sources \label{corepara}} \tablehead{
\colhead{Name} &  \colhead{RA Offset\tablenotemark{a}} &  \colhead{DEC Offset\tablenotemark{a}}  & \colhead{ I$_{peak}$\tablenotemark{b} } & \colhead{ S$_{\nu}$\tablenotemark{c} } & \colhead{a\tablenotemark{d}} & \colhead{b\tablenotemark{d}} &
\colhead{P.A.\tablenotemark{d}} & \colhead{R\tablenotemark{d}} & \colhead{M\tablenotemark{e}} & \colhead{n\tablenotemark{e}} & \colhead{Radio source\tablenotemark{f}}  \\
\colhead{}  & \colhead{($\arcsec$)}& \colhead{($\arcsec$)} &
\colhead{(10$^{-2}$ Jy~beam$^{-1}$)} & \colhead{(10$^{-2}$ Jy)} &
\colhead{(arcsec)} & \colhead{(arcsec)} &
\colhead{(degree)} & \colhead{($10^3$ AU)}  & \colhead{(M$_{\sun}$)} & \colhead{($10^6$ cm$^{-3}$)} } \startdata
MM1       &       -4.31(0.07)	   &    7.04(0.06)                &  1.33(0.09)  &   9.50     &    2.61(0.21)  &  1.53(0.11)   &   -53.8(6.3)  & 10.4     & 30.8(60.7)  & 1.0(1.9)    &   \\
MM2       &       -4.08(0.04)	   &    4.40(0.05)                &  0.58(0.16)  &   1.24     &    1.06(0.14)  &  0.62(0.06)   &   -24.7(26.9) & 4.2      &  4.0(7.9)   & 1.9(3.8)    &    \\
MM3       &       -0.33(0.02)	   &    4.77(0.04)                &  1.56(0.14)  &   4.16     &    1.58(0.10)  &  0.43(0.02)   &     0.4(4.1)  & 4.3      & 13.5(26.6)  & 6.1(12.1)   &     \\
MM4       &       -1.86(0.18)	   &    3.13(0.11)                &  2.56(0.14)  &   22.84    &    2.58(1.05)  &  1.05(0.21)   &   -71.3(9.5)  & 8.6      & 74.0(23.4)  & 4.2(1.3)    &  E,Gau \\
          &       -1.65(0.04)	   &    3.38(0.03)                &  9.26(0.31)  &            &                &               &               &          &             &             &   E,Point  \\
MM5       &       -1.12(0.11)	   &    1.86(0.07)                &  1.44(0.14)  &   4.60     &    1.53(0.24)  &  0.95(0.16)   &   -86.8(15.8) & 6.3      & 14.9(29.4)  & 2.2(4.3)    &   \\
MM6       &        0.01(0.03)	   &    0.05(0.03)                &  3.44(0.30)  &   9.03     &    1.17(0.08)  &  0.78(0.04)   &    -5.9(19.0) & 5.0      & 29.3(57.7)  & 8.5(16.8)   &    \\
          &       -0.10(0.43)	   &    0.12(0.37)                &  0.46(0.47)  &            &                &               &               &          &             &             &  \\
MM7       &        0.27(0.03)	   &   -2.31(0.03)                &  3.43(0.18)  &   5.47     &    0.75(0.06)  &  0.43(0.03)   &    -4.5(93.4) & 3.0      & 17.7(5.6)   & 24.6(7.8)   &  G  \\
MM8       &        1.08(0.03)	   &   -4.64(0.13)                &  3.91(0.14)  &   26.72    &    2.70(0.52)  &  0.67(0.03)   &     4.6(2.6)  & 7.0      & 86.6(27.4)  &  9.1(2.9)   &  F,Gau  \\
          &        1.26(0.03)	   &   -4.50(0.02)                &  8.08(0.30)  &            &                &               &               &          &             &             &  F,Point  \\
MM9       &        3.50(0.08)	   &   -6.37(0.04)                &  2.70(0.18)  &   4.21     &    0.71(0.09)  &  0.51(0.06)   &   -70.9(15.8) & 3.1      & 13.6(26.9)  & 15.9(31.4)  & \\
MM10      &        2.16(0.11)	   &   -6.50(0.13)                &  1.86(0.12)  &   6.72     &    1.71(0.23)  &  0.93(0.15)   &   -33.0(17.8) & 6.6      & 21.8(42.9)  &  2.8(5.4)   &  \\
MM11      &        2.17(0.07)	   &   -8.59(0.16)                &  1.97(0.12)  &   24.94    &    2.14(0.22)  &  1.27(0.12)   &   -28.8(13.1) & 8.6      & 80.8(25.6)  &  4.6(1.5)   &  D,Gau \\
          &         2.11(0.04)	   &   -8.05(0.03)              &  4.42(0.28)  &            &                &               &               &          &             &             &  D,Point   \\
MM12      &        3.66(0.08)	   &  -11.38(0.17)                &  0.60(0.07)  &   3.16     &    2.14(0.37)  &  1.24(0.49)   &    -1.2(25.1) & 8.5      & 10.2(20.2)  &  0.6(1.2)   &
\enddata
\tablenotetext{a}{Offsets with respect to the phase center R.A.(J2000)~=~18$^{\rm h}$06$^{\rm m}$14.78$^{\rm s}$ and DEC.(J2000)~=~$-20\arcdeg31\arcmin34.9\arcsec$.}
\tablenotetext{b}{peak flux density}
\tablenotetext{c}{Total flux density from 2-D Gaussian fit. For MM4, MM6, MM8 and MM11, the total flux is the sum of the fluxes of the Gaussian component and the point source component. }
\tablenotetext{d}{a, b, P.A., are deconvolved FWHM major size, minor size and position angle from 2D Gaussian fits. The effective radiu (R) is defined as $R=\sqrt{ab}$ }
\tablenotetext{e}{The values in parentheses are derived by assuming a dust temperature T$_{d}$ of 100 K for MM4, MM7, MM8 and MM11; and T$_{d}$ of 20 K for other cores. The values outside parentheses are derived by assuming T$_{d}$ of 35 K for all cores. }
\tablenotetext{e}{``Gau" means Gaussian component; ``Point" means point-like component.}
\end{deluxetable}
\end{longrotatetable}

\clearpage

\subsection{Line emission}

\subsubsection{JCMT C$^{18}$O (3-2) line and virial analysis of the G9.62 clump}

\begin{figure}[tbh!]
\centering
\includegraphics[angle=0,scale=0.3]{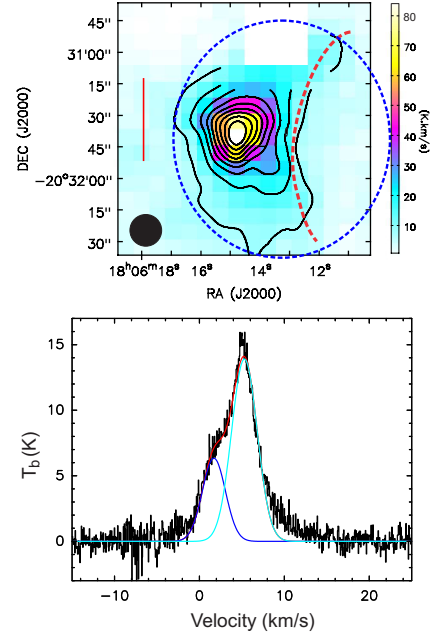}
\caption{Upper panel: Integrated intensity (from -2 to 10 km~s$^{-1}$) map of C$^{18}$O (3-2) line. The contours are from 10\% to 90\% in steps of the 10\% of the peak value (82 K~km~s$^{-1}$). The rms level in the map is about 0.7 K km~s$^{-1}$. The beam is shown in filled black circle. The vertical red line represents a spatial scale of 1 pc. The western edge of the gas clump seems to be bent as indicated by the red dashed line. The blue dashed circle outlines the unperturbed natal clump. Lower panel: the C$^{18}$O (3-2) spectrum at the peak position. The blue and red lines are gaussian fits.\label{C18O}}
\end{figure}

Figure \ref{C18O} shows the integrated intensity map of C$^{18}$O (3-2) and its spectrum at pixel of the emission peak. The C$^{18}$O (3-2) emission shows a single gas clump with an effective radius of $\sim$0.9 pc. Interestingly, the western edge of the gas clump seems to be bent as indicated by the red dashed line, indicating that the gas clump may be compressed by H{\sc ii} region to the west. The blue dashed circle may outline the unperturbed natal clump. The spectrum can be well fitted with two gaussian components. The blueshifted component with a linewidth of $\sim$3.2$\pm$0.1 km~s$^{-1}$ peaks at 1.7 km~s$^{-1}$, while the redshifted one with a linewidth of 3.5$\pm$0.1 km~s$^{-1}$ peaks at 5.3 km~s$^{-1}$. The two velocity components originate from the two sub-clumps as revealed in the JCMT/SCUBA 450 $\micron$ continuum image. Previous SMA molecular line observations have indicated that the northern sub-clump has a systemic velocity around 2-3 km~s$^{-1}$, while the southern sub-clump has a systemic velocity around 5-6 km~s$^{-1}$ \citep{liu11}, which are consistent with the C$^{18}$O (3-2) results. To examine the gravitational stability of the G9.62 clump, we calculated its virial mass. The virial mass considering turbulent support can be derived as $\frac{M_{vir}}{M_{\sun}}=210(\frac{R}{pc})(\frac{\Delta V}{km~s^{-1}})^2$ \citep{mac88,zhang15}, where $\Delta V$ is the linewidth of C$^{18}$O (3-2). With a radius of $\sim$0.25 pc ($\sim$10$\arcsec$; derived from SCUBA 450 $\micron$ map as shown in Table \ref{fluxpara} in Appendix), the virial masses for the northern and southern sub-clumps are $\sim$540$\pm$81 and $\sim$640$\pm$96 M$_{\sun}$, respectively.

The uncertainties of the virial masses are mainly determined by the radii. The uncertainties of the radii from 2-D Gaussian fits are $\sim$15\%. The uncertainties of virial masses caused by linewidths measurements are negligible because the errors from Gaussian fits are small and C$^{18}$O (3-2) line emission is usually optically thin. The optical depth of C$^{18}$O (3-2) line may cause overestimation of the linewidth if its emission is optically thick. Assuming an excitation temperature of 35 K, the same as dust temperature, the optical depths of blueshifted component and redshifted component are $\sim$0.3 and $\sim$0.7, respectively, indicating that C$^{18}$O (3-2) line emission is optically thin. The total virial mass of the G9.62 clump is $\sim1200\pm180$ M$_{\sun}$, much smaller than the mass (2800$\pm$200 M$_{\sun}$; see Figure \ref{SED}) estimated from dust continuum, indicating that the G9.62 clump will continue to collapse. The contraction of the G9.62 clump could be further enhanced if considering the external compression from the western H{\sc ii} regions.

\subsubsection{Molecular lines from ALMA observations}

\begin{figure*}[tbh!]
\centering
\includegraphics[angle=0,scale=0.8]{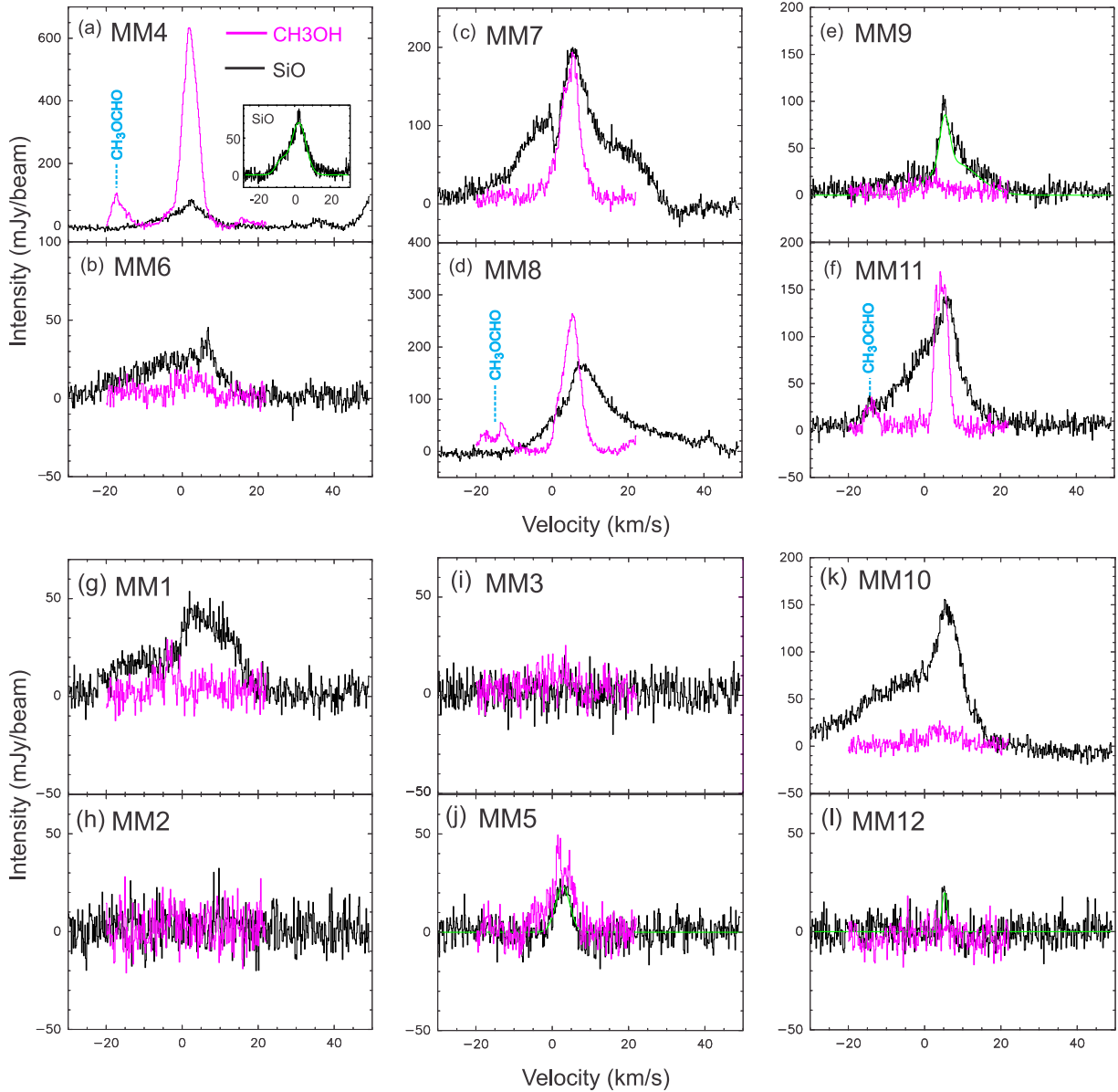}
\caption{Spectra of SiO (5-4) (black) and CH$_{3}$OH v$_{t}$=1 ($6_{1,5}-7_{2,6}$) (pink). The spectra were core-averaged. The upper panels (a-f) show the spectra at the 6 densest cores, while the lower panels (g-l) show the spectra at the other 6 less dense cores. The green lines in panels (a, e, j ,l) are Gaussian fits of the SiO (5-4) spectra. The CH$_{3}$OCHO ($17_{4,13}-16_{4,12}$) emission in panels (a,d,f) were marked with blue dashed lines.\label{chem}}
\end{figure*}

\begin{figure}[tbh!]
\centering
\includegraphics[angle=0,scale=0.4]{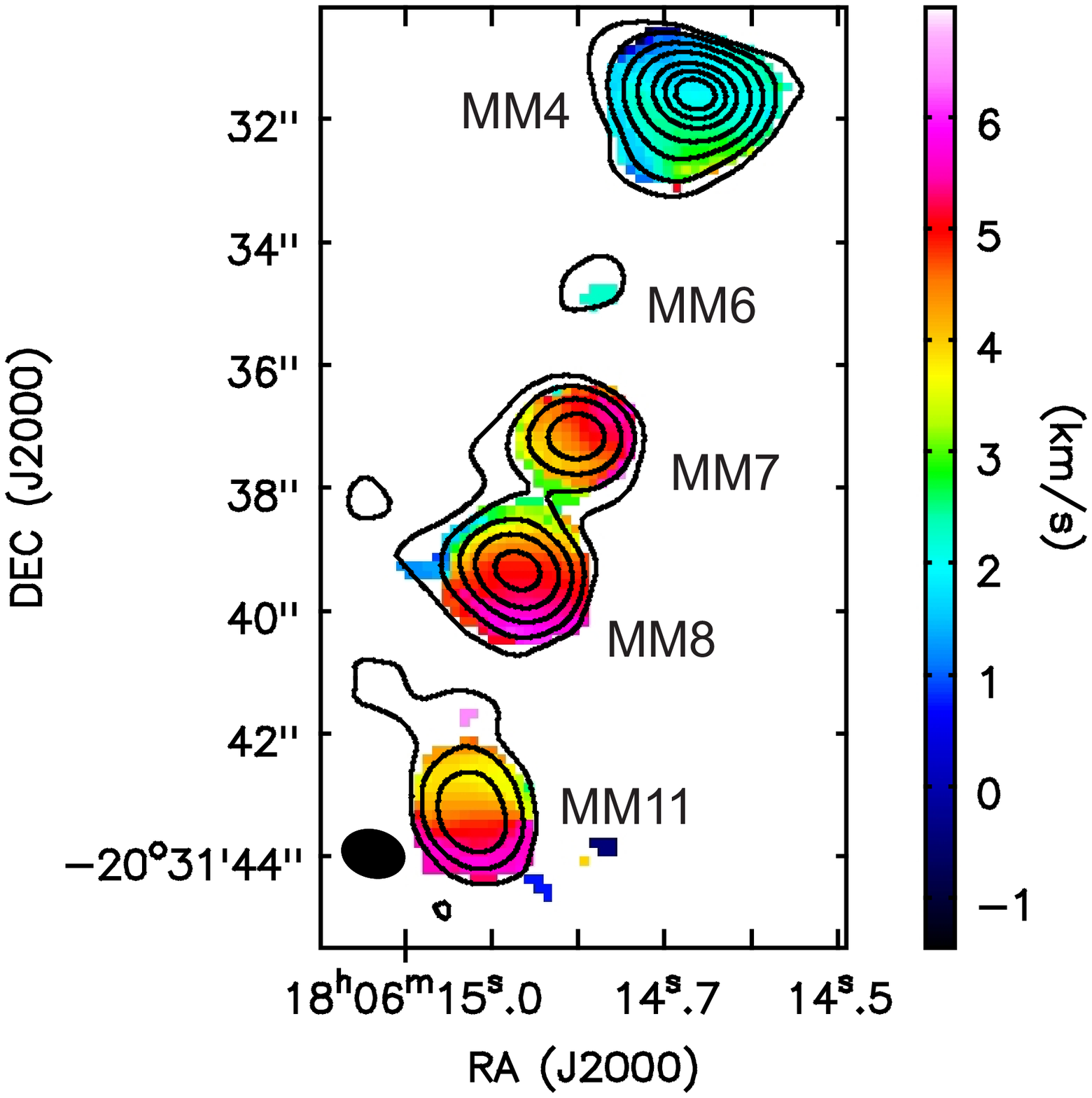}
\caption{The integrated intensity (from -6 to 17 km~s$^{-1}$) map of CH$_{3}$OH v$_{t}$=1 ($6_{1,5}-7_{2,6}$) is shown in contours overlaid on its Moment 1 color image. The contour levels are [0.03,0.05,0.1,0.2,0.4,0.6,0.8]$\times$6.82 Jy~beam$^{-1}$~km~s$^{-1}$. The rms level of the integrated intensity map is about 0.06Jy~beam$^{-1}$~km~s$^{-1}$. \label{CH3OH}}
\end{figure}

The upper panels (a-f) of Figure \ref{chem} show the spectra of SiO (5-4) and CH$_{3}$OH v$_{t}$=1 ($6_{1,5}-7_{2,6}$) averaged over the core regions of the six brightest cores. MM7 (radio source ``G") and MM8 (``radio source F") show strong line wings in SiO (5-4), suggesting the existence of outflows. MM6 shows a broad blueshifted line wing with terminal velocity up to $\sim$20 km~s$^{-1}$. MM9 and MM11 also show line wings in SiO (5-4). However, the line wings in SiO (5-4) toward MM9 and MM11 may be caused by the contamination of the widespread outflows associated with MM8 (see section 3.4). Strong CH$_{3}$OH v$_{t}$=1 ($6_{1,5}-7_{2,6}$) emission was detected toward four dense cores, MM4, MM7, MM8 and MM11. We fitted their spectra with gaussian profiles and present their velocities and line widths in Table \ref{linepara}. MM4 has a systemic velocity of $\sim$2 km~s$^{-1}$, while the other three cores have a systemic velocity of $\sim$5 km~s$^{-1}$, which are consistent with the C$^{18}$O results as well as previous SMA observations \citep{liu11}. MM4 (radio source ``E") shows the strongest CH$_{3}$OH v$_{t}$=1 ($6_{1,5}-7_{2,6}$) emission in the G9.62 clump.  MM6 and MM9 do not show CH$_{3}$OH v$_{t}$=1 ($6_{1,5}-7_{2,6}$) emission.

The lower panels (g-l) of Figure \ref{chem} show the spectra of SiO (5-4) and CH$_{3}$OH v$_{t}$=1 ($6_{1,5}-7_{2,6}$) toward the other six less bright cores. MM1 shows broad SiO (5-4) emission but no CH$_{3}$OH v$_{t}$=1 ($6_{1,5}-7_{2,6}$) emission. No SiO (5-4) or CH$_{3}$OH v$_{t}$=1 ($6_{1,5}-7_{2,6}$) emission is detected toward MM2 and MM3. MM5 is located to the south of the UC H{\sc ii} region MM4/E but shows much narrower SiO (5-4) emission than MM4/E. Its CH$_{3}$OH v$_{t}$=1 ($6_{1,5}-7_{2,6}$) shows double-peaked profile with blueshifted one stronger than the red one, a typical ``blue profile", indicating that the core may be in collapse \citep{zhou93}. No CH$_{3}$OH v$_{t}$=1 ($6_{1,5}-7_{2,6}$) emission is detected toward MM10. MM10 shows very broad SiO (5-4) emission with blueshifted line wings. However, since MM10 is is located at the center of the blueshifted outflow lobe (see section 3.4), the broad SiO (5-4) emission at MM10 is very likely caused by the outflows.

Figure \ref{CH3OH} presents the integrated intensity (from -6 to 17 km~s$^{-1}$) map of CH$_{3}$OH v$_{t}$=1 ($6_{1,5}-7_{2,6}$) in contours overlaid on its Moment 1 color image (or intensity weighted velocity map). The CH$_{3}$OH v$_{t}$=1 ($6_{1,5}-7_{2,6}$) emission is mainly distributed in cores MM4, MM7, MM8 and MM11. The other cores have no obvious emission above 3$\sigma$ noise level. The Moment 1 map of these four cores show clear velocity gradients, suggesting rotational motions. The discussions of the Moment 1 map are beyond the scope of this paper and will be presented in another work. The integrated intensity maps of SiO (5-4) are presented in Figure \ref{mom0} and will be discussed in section 3.3.

\begin{deluxetable}{ccc}
\centering
\scriptsize
\tablecolumns{3} \tablewidth{0pc}
\tablecaption{Line parameters \label{linepara}} \tablehead{
\colhead{Name}  & \colhead{V$_{LSR}$} & \colhead{FWHM}  \\
\colhead{}  & \colhead{(km~s$^{-1}$)} & \colhead{(km~s$^{-1}$)} }
\startdata
\multicolumn{3}{c}{CH$_{3}$OH v$_{t}$=1 ($6_{1,5}-7_{2,6}$)}\\
\cline{1-3}
MM4/E&2.13(0.01)&4.70(0.02)\\
MM7/G&4.89(0.03)&5.87(0.07)\\
MM8/F&5.11(0.01)&5.51(0.03)\\
MM11/D&4.51(0.02)&3.76(0.04)\\
\cline{1-3}
\multicolumn{3}{c}{SiO (5-4)}\\
\cline{1-3}
MM4/E\tablenotemark{a} &2.33(0.12) & 9.48(0.29)\\
MM5   &2.75(0.15) & 4.40(0.31)\\
MM9\tablenotemark{a}   &5.41(0.09) & 3.03(0.26) \\
MM12  &5.25(0.11) & 1.66(0.34)\\
\cline{1-3}
\multicolumn{3}{c}{[Ne {\sc ii}]}\\
\cline{1-3}
A&0.84(0.14)&15.61(0.30)\\
B&1.26(0.04)&12.53(0.09)\\
C-e&-2.31(0.35)&14.19(0.86)\\
C-w&-5.67(0.24)&9.31(0.66)\\
\enddata
\tablenotetext{a}{Core emission close to the systemic velocity}
\end{deluxetable}

\subsubsection{12.8 $\micron$ [Ne {\sc ii}] line}

\begin{figure*}[tbh!]
\centering
\includegraphics[angle=90,scale=0.6]{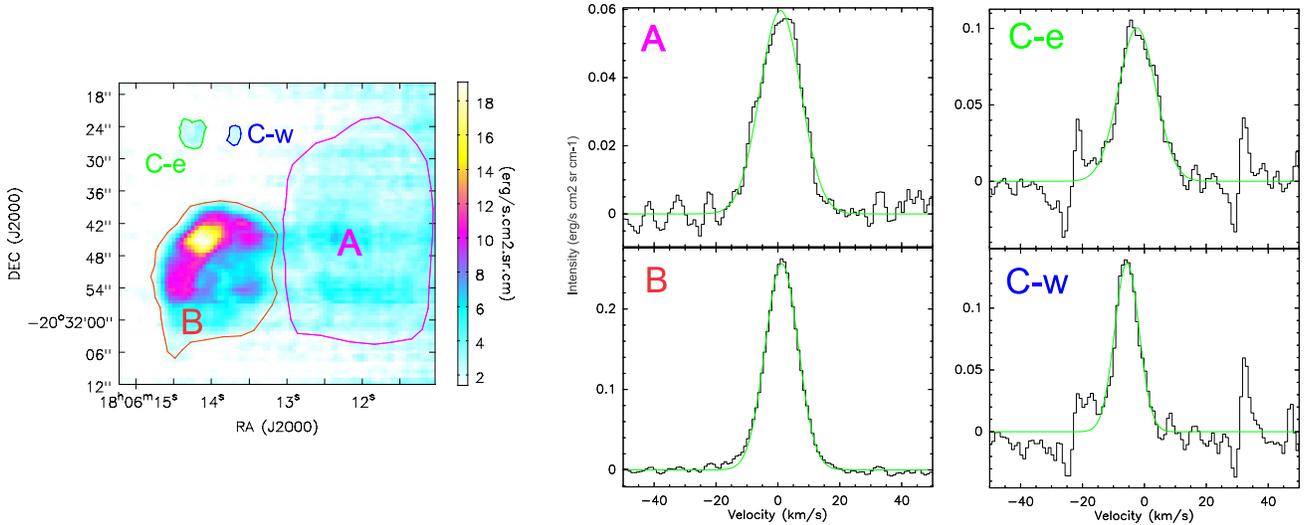}
\caption{The left panel shows the integrated intensity map of [Ne {\sc ii}] line. The boundary of each component (``A", ``B", ``C-e" \& ``C-w") roughly corresponds to the contour of 10\% of the peak value. The right panels show the averaged spectra of each component. The green lines are Gaussian fits.\label{NeII} }
\end{figure*}

In contrast to radio recombination lines (RRLs) whose high thermal line widths ($\sim$20 km~s$^{-1}$ for a electron temperature of 10$^{4}$ K) are comparable to the velocities of bulk flows in H{\sc ii} regions, Mid-IR fine-structure lines (e.g., 12.8 $\micron$ [Ne {\sc ii}] line) have much smaller thermal line widths because they are emitted by heavy elements. Therefore, high spectral and
spatial resolution maps of 12.8 $\micron$ [Ne {\sc ii}] fine-structure lines are much more suitable to study the kinematic patterns in the ionized gas than RRLs and radio continuum emission \citep{zhu05,zhu08}.

The left panel of Figure \ref{NeII} shows the integrated intensity map of 12.8 $\micron$ [Ne {\sc ii}] line emission. The [Ne {\sc ii}] line emission is only detected toward radio sources ``A", ``B" and ``C" \citep{gar93}. Radio source ``C" has two components, ``C-e" and ``C-w". The boundaries of the four components are marked with color outlines. We did not detect [Ne {\sc ii}] toward the dense cores in the G9.62 clump including those that show radio continuum presumably due to the high extinction there. Radio source ``A" is much larger and more diffuse than ``B" and ``C", indicating that ``A" is older. From west to east, sequential high-mass star formation (from old to young: ``A"$\rightarrow$``B"\&``C"$\rightarrow$G9.62 clump) is taking place at pc scale in the G9.62+0.19 complex as also suggested by \cite{hof94}.

The right panels of Figure \ref{NeII} present the averaged spectra of each component. The velocities and linewidths of the spectra are summarized in Table \ref{linepara}. The radio sources ``A", ``B" and ``C" show different velocities from the millimeter sources detected by ALMA. In general, their velocities are more blueshifted when compared to the dense cores in the G9.62 clump. The PDR of radio source ``B" traced by 8 $\micron$ emission (see Figure 1) shows very dense contours in the region adjacent to the G9.62 clump, strongly suggesting that radio source ``B" is interacting with the G9.62 clump. However, the velocity of radio source ``B" is blueshifted by $\sim$5 km~s$^{-1}$ relative to the G9.62 clump. Similarly, radio source ``C-e" is close to the northern dense cores (e.g., MM4) but its velocity is blueshifted by $\sim$4.5 km~s$^{-1}$. Such a velocity difference between ionized gas and molecular gas may indicate that the molecular gas was compressed by the shock front induced by the ionization front of the H{\sc ii} regions. The shock velocity could be as high as $\sim$10 km~s$^{-1}$ if we consider a moderate inclination angle \citep[see also][]{hof94}. The [Ne {\sc ii}] line will be modelled and analyzed in another paper (Zhu et al. 2017, in preparation).

\subsection{Molecular outflows}

\begin{figure}[tbh!]
\centering
\includegraphics[angle=0,scale=0.5]{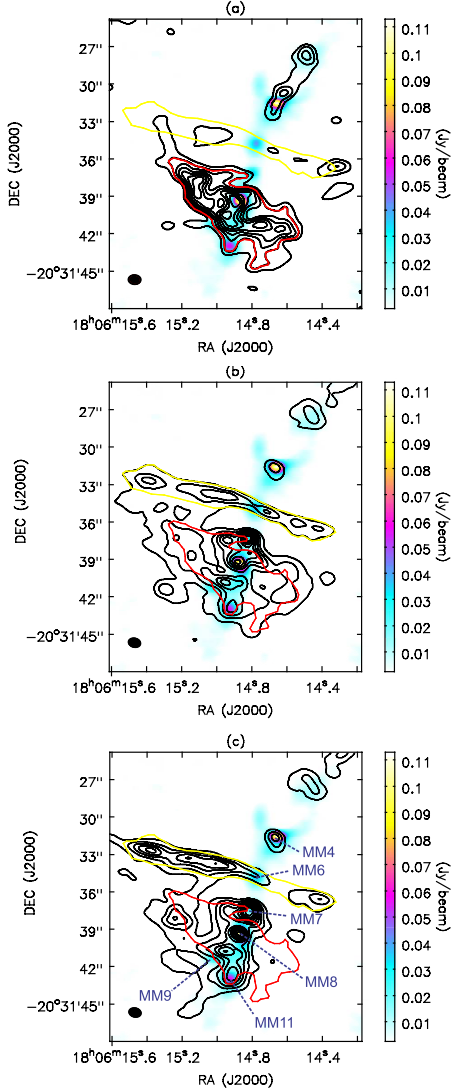}
\caption{The 1.3 mm continuum is shown as color images in three panels. (a). The integrated intensity (from -70 to 90 km~s$^{-1}$) of CO (2-1) is shown in contours. The contours are [0.1, 0.2, 0.3, 0.4, 0.5, 0.6, 0.7, 0.8, 0.9]$\times$29 Jy/beam~km~s$^{-1}$. The red outline corresponds to the 20\% contour, which marks the main CO emission region. (b). The integrated intensity (from -29 to 48 km~s$^{-1}$) of SiO (5-4) is shown in contours. The contour levels are [0.1, 0.2, 0.3, 0.4, 0.5, 0.6, 0.7, 0.8, 0.9]$\times$7.07 Jy/beam~km~s$^{-1}$. The yellow outline corresponds to the 10\% contour, which marks the elongated bar-like structure in SiO emission. (c). The integrated intensity (from -1 to 10 km~s$^{-1}$) of SiO (5-4) is shown in contours.
The contour levels are [0.1, 0.2, 0.3, 0.4, 0.5, 0.6, 0.7, 0.8, 0.9]$\times$2.72 Jy/beam~km~s$^{-1}$. The names of the six densest cores are labeled. \label{mom0}}
\end{figure}

\begin{figure*}[tbh!]
\centering
\includegraphics[angle=90,scale=0.6]{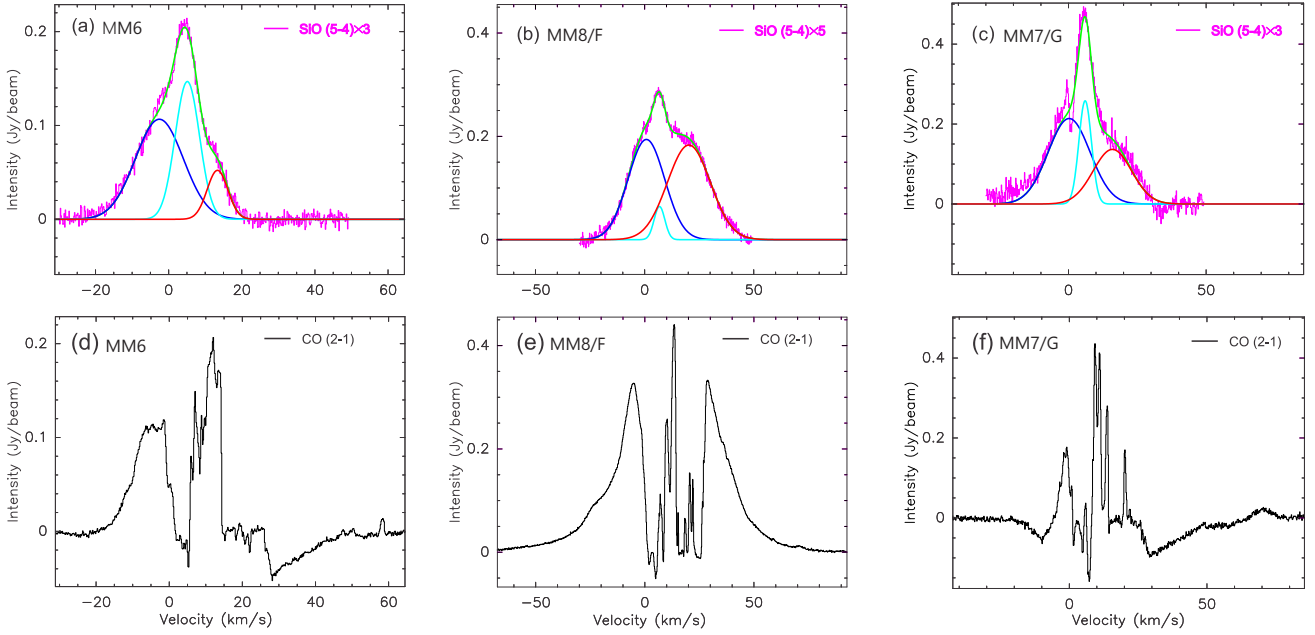}
\caption{Upper panels: Spectra of SiO (5-4) are shown in magenta. The red, blue and cyan lines are Gaussian fits. The green lines are the sum of the Guassian fits. Lower panels: Spectra of CO (2-1); the absorption features in CO spectra are caused by self-absorption or foreground clouds along the line of sight. Spectra of CO (2-1) and SiO (5-4) for MM6 (a \& d) are averaged over the yellow outlined area in Figure \ref{mom0}. Spectra of CO (2-1) and SiO (5-4) for MM8/F (b \& e) are averaged over the red outlined area in Figure \ref{mom0}. Spectra of CO (3-2) and SiO (5-4) of MM7/G (c \& f) are averaged over 2$\arcsec$ area centered on MM7 to cover its outflow emission region.\label{spectra}}
\end{figure*}

\begin{figure}[tbh!]
\centering
\includegraphics[angle=0,scale=0.4]{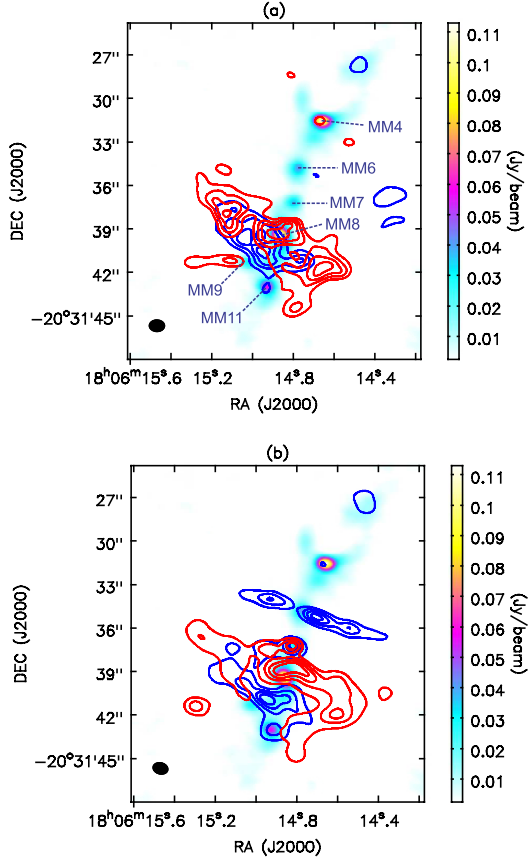}
\caption{(a). CO outflows.
The redshifted emission is integrated in [29km/s, 100km/s] interval. The contours are [0.2, 0.4, 0.6, 0.8]$\times$13.1 Jy/beam~km~s$^{-1}$; The blueshifted emission is integrated in [-70km/s, -6km/s] interval. The contours are [0.2, 0.4, 0.6, 0.8]$\times$21.9 Jy/beam~km~s$^{-1}$. The names of the six densest cores are labeled.
(b). SiO outflows.
The redshifted emission is integrated in [12km/s, 48km/s] interval. The contours are [0.2, 0.4, 0.6, 0.8]$\times$2.89 Jy/beam~km~s$^{-1}$; The blueshifted emission is integrated in [-29km/s, -2km/s] interval. The contours are [0.2, 0.4, 0.6, 0.8]$\times$2.02 Jy/beam~km~s$^{-1}$.\label{outflow}}
\end{figure}

\begin{deluxetable}{cccccc}
\centering
\scriptsize
\tablecolumns{6} \tablewidth{0pc}\setlength{\tabcolsep}{0.05in}
\tablecaption{Outflow parameters \label{outflowpara}} \tablehead{
\colhead{Lobes} & \colhead{$\int S_{\nu}dV$} & \colhead{M} & \colhead{P} & \colhead{E} & \colhead{V$_{char}$\tablenotemark{a}} \\
\colhead{} & \colhead{(Jy~km~s$^{-1}$)} & \colhead{(M$_{\sun}$)} & \colhead{(M$_{\sun}~km~s^{-1}$)} & \colhead{($10^{46}$erg)} & \colhead{(km~s$^{-1}$)}}
\startdata
\multicolumn{6}{c}{SiO outflows of MM6\tablenotemark{b}}\\
\cline{1-6}
red  & 4.7 & 14(0.4) & 340(8.5) & 9.0(0.2) & 24.4 \\
blue & 13.1 & 39(1.0) & 1300(32.5) & 50.7(1.3) & 33.2 \\
\cline{1-6}
\multicolumn{6}{c}{CO outflows of MM6\tablenotemark{c}}\\
\cline{1-6}
blue & 38.8& 0.1(0.9) & 3.4(17) & 0.12(0.24) & 33.0(19) \\
\cline{1-6}
\multicolumn{6}{c}{SiO outflows of MM7/G\tablenotemark{d}}\\
\cline{1-6}
red &2.5& 5.2(0.5) & 62(6.2) & 0.9(0.1) & 12.0 \\
blue&3.3& 6.7(0.7) & 90(0.9) & 1.6(0.2) & 13.5 \\
\cline{1-6}
\multicolumn{6}{c}{SiO outflows of MM8/F\tablenotemark{e}}\\
\cline{1-6}
red  &35.7& 73(7.3) & 164(16.4) & 44.8(4.5) & 22.6\\
blue &21.0& 43(4.3) & 73(7.3) & 14.3(1.4) & 16.8 \\
\cline{1-6}
\multicolumn{6}{c}{CO outflows of MM8/F\tablenotemark{f}}\\
\cline{1-6}
red & 290.0 & 0.8(7.2) & 30(150) & 1.4(2.8) & 38(21)\\
blue & 282.1& 0.8(7.2) & 24(120) & 1.0(2.0) & 31(17)\\
\enddata
\tablenotetext{a}{V$_{char}$=P/M}
\tablenotetext{b}{Assuming an inclination angle of 70$\arcdeg$, SiO excitation temperature of 30 K and SiO abundance of 1.4$\times10^{-10}$.}
\tablenotetext{c}{Assuming an inclination angle of 70$\arcdeg$, CO excitation temperature of 30 K and CO abundance of 1.0$\times10^{-4}$.}
\tablenotetext{d}{Assuming an inclination angle of 0$\arcdeg$, SiO excitation temperature of 40 K and SiO abundance of 2.1$\times10^{-10}$.}
\tablenotetext{e}{Assuming an inclination angle of 45$\arcdeg$, SiO excitation temperature of 40 K and SiO abundance of 2.1$\times10^{-10}$.}
\tablenotetext{f}{Assuming an inclination angle of 45$\arcdeg$, CO excitation temperature of 30 K and CO abundance of 1.0$\times10^{-4}$.}
\tablenotetext{g}{The values in parentheses are derived by assuming SiO abundance enhancement factors of 10 and 40 for MM7/MM8 and MM6, respectively; and applying optical depth correction on CO (2-1) emission.}
\end{deluxetable}

CO and SiO are good outflow tracers. Figure \ref{mom0} shows the integrated intensity maps of CO (2-1) and SiO (5-4). Both CO (2-1) and SiO (5-4) line emission distributions are very clumpy. As marked by the red outline, the CO (2-1) line emission is mainly distributed in the south of the G9.62 clump (left panel). An elongated structure is seen in the north. There is a very long ($\sim$18$\arcsec$) bar-like structure in SiO (5-4) line emission, as marked by the yellow outline in the middle panel. This elongated structure suggests a well collimated outflow originating from MM6. In the right panel, we only integrated the intensity from -1 to 10 km~s$^{-1}$ for SiO (5-4). In this velocity interval, the SiO (5-4) line emission is generally associated with dense cores. However, the bar-like structure is still obvious in such low-velocity emission, indicating that the outflow associated with MM6 may have a very large inclination angle with respect to the line of sight. Therefore, the majority of outflow emission shows slow velocity and the outflow lobes are nearly parallel to the plane of the sky.

Figure \ref{spectra} presents the averaged spectra of CO (2-1) and SiO (5-4) in the yellow outlined region, red outlined region and dense core MM7/G. The SiO spectra show high velocity wings in all the three regions. The outflow emission and ambient gas emission in SiO (5-4) spectra can be well separated with three Gaussian lines. The ambient low-velocity gas emission in SiO (5-4) has a systemic velocity $\sim$5-7 km~s$^{-1}$ and velocity dispersion of $\sim$5-8 km~s$^{-1}$. CO (2-1) spectra are quite different from SiO (5-4). Only blueshifted high velocity CO (2-1) emission is detected in the outflow associated with MM6 as shown in panel (a). As shown in panel (b), the CO (2-1) emission in the southern region shows extremely high-velocity wings, indicating that the CO emission there is dominated by outflows. No line wings in CO (2-1) are detected toward dense core MM7/G.

Figure \ref{outflow} shows the integrated intensity maps of high-velocity emission of CO (2-1) and SiO (5-4). The high-velocity CO (2-1) emission reveals a clumpy outflow in the southern region. The red and blue lobes of this CO outflow are roughly overlapped. This widespread CO outflow should originate from the hot core MM8/F as also suggested by \cite{hof01,liu11}. The outflow from the hot core MM8/F is very energetic with a maximum flow velocity up to $\sim$100 km~s$^{-1}$ in CO (2-1) emission assuming an inclination angle of 45$\arcdeg$. Besides the outflow from MM8/F, the high-velocity SiO (5-4) emission also clearly reveals the outflows originating from MM6 and MM7/G. As mentioned above, the outflow associated with MM6 is very collimated and dominated by blueshifted emission.

Following \cite{qiu07,qiu09,wang11,zhang15,liu11,liu15a,liu16a}, we calculated the outflow masses (M), momentum (P) and kinetic energy (E) in each velocity channel assuming optically thin emission and present their sums in Table \ref{outflowpara}. The integrated fluxes of SiO (5-4) and CO (2-1) line emission used in calculating outflow masses are presented in the second column of Table \ref{outflowpara}. As mentioned above, the outflow associated with MM6 may have a very large inclination angle with respect to the line of sight. We assume an inclination angle of 70$\arcdeg$ to calculate the outflow parameters for MM6. The blue and red outflow lobes of MM7/G are overlapped along the line of sight, indicating that its outflow inclination angle is very small. We take an inclination angle of 0$\arcdeg$ to calculate the outflow parameters for MM7/G. Since the blue and red outflow lobes of MM8/F are well separated and have very high flow velocities, neither the very low (pole-on) inclination nor the very high (edge-on) inclination is likely. Therefore, we take 45$\arcdeg$ as the inclination angle for MM8/F. The inclination angles do not affect the determination of outflow masses. The inclination angle ($\theta$) corrections on flow velocities and outflow momentum are by a factor of $1/cos(\theta)$. For outflow energy, the correction factors are $1/cos^2(\theta)$.

\cite{gern14} found that the median SiO abundances in HMPOs and hot cores are 1.4$\times10^{-10}$ (for T=30 K) and 2.1$\times10^{-10}$ (for T=40 K) from a single-dish IRAM 30-m suvey of 59 high-mass star forming regions. We adopt an excitation temperature of 30 K and SiO abundance of 1.4$\times10^{-10}$ for the outflow of the HMPO MM6. An excitation temperature of 40 K and SiO abundance of 2.1$\times10^{-10}$ were adopted for outflows of hot cores MM7/G and MM8/F. The total outflow masses of MM6, MM7/G and MM8/F derived from SiO (5-4) emission are 53, 13 and 116 M$_{\sun}$, respectively. A typical abundance of 10$^{-4}$ for CO is used to derive the CO outflow parameters. The excitation temperature for CO outflow emission is taken as the same of SiO. The total outflow mass of MM8/F derived from CO (2-1) is $\sim$1.9 M$_{\sun}$. The outflow mass of MM6 blue lobe derived from CO (2-1) is $\sim$0.1 M$_{\sun}$.

The uncertainties of the outflow parameters caused by integrated flux measurements are negligible because of the high S/N levels in outflow emission. The main uncertainties of outflow parameters are caused by excitation temperatures, abundances and optical depths. If we adopt an alternative excitation temperature of 20 K or 40 K for the outflow of MM6, the outflow parameters will increase or decrease by $\sim$50\%, respectively. The outflow parameters of MM7/G and MM8/F will become 50\% larger if we adopt an excitation temperature of 30 K instead of 40 K. SiO high velocity emission is usually optically thin due to its low abundance while CO high velocity emission is usually optically thick \citep{qiu07,qiu09}. People can use $^{12}$CO/$^{13}$CO line ratios to correct for $^{12}$CO opacity as a function of velocity in previous outflow studies \citep{qiu09,Zhang16}.  Unfortunately, we did not observe $^{13}$CO line. In the HH 46/47 molecular outflow, the mass, momentum, and kinetic energy derived from CO (2-1) were underestimated by factors of about 9, 5, and 2, respectively, if assuming optically thin emission \citep{Zhang16}. We applied the same correction factors to the CO outflow parameters and list the new values in parentheses in Table \ref{outflowpara}. These corrections need to be tested by future $^{13}$CO line observations. However, after correction of optical depth, the total outflow mass (14.4 M$_{\sun}$) derived from CO (2-1) for MM8/F is very consistent with previous studies with SO, CS or HCN lines \citep{liu11}, indicating that the corrections are reasonable.
The abundance of CO in outflow region usually not changes too much. However, SiO abundance is usually greatly enhanced in outflow regions, which will lead to an overestimation of outflow parameters. The outflow parameters (M, P, E) calculated from SiO (5-4) are orders of magnitudes larger than the outflow parameters derived from CO (2-1). For example, the total outflow mass of MM8/F derived from CO (2-1) (1.9 M$_{\sun}$) is about 60 times smaller than the value derived from SiO (5-4).  Even though CO emission is corrected with optical depth, the outflow mass estimated from CO (2-1) is still 8 times smaller than the value derived from SiO (5-4), indicating that the SiO abundance is underestimated at least by a factor of $\sim$10 in the outflows of MM8/F. The optical depth corrected outflow mass for the blueshifted lobe of MM6 derived from CO (2-1) is about 40 times smaller than the value derived from SiO (5-4), suggesting that the SiO abundance might be underestimated by a factor of $\sim$40 in the outflows of MM6. Therefore, the SiO abundance in outflow regions of MM8/F and MM6 should be enhanced by a factor of $\sim$10 and $\sim$40, respectively, as compared with clump averaged values in high-mass star forming regions \citep{gern14}. In Table \ref{outflowpara}, we also list the outflow parameters estimated from SiO (5-4) by applying abundance enhancement factors of 10 and 40 for MM7/MM8 and MM6, respectively. Then the total outflow masses of MM6, MM7/G and MM8/F derived from SiO (5-4) are 1.4, 1.2 and 11.6 M$_{\sun}$, respectively. The outflow mass of MM8/F derived from SiO (5-4) is now consistent with outflow mass derived from CO (2-1) by applying optical depth correction. In below analysis, we use the outflow parameters derived from SiO (5-4) considering abundance enhancement for MM6 and MM7/G to derive their outflow mass loss rates and accretion rates. For MM8/F, we use outflow parameters derived from CO (2-1) by applying optical depth correction.

Assuming an inclination angle of 70$\arcdeg$, the length (L) and maximum flow velocity (V$_{max}$) of MM6 outflow lobes are $\sim$0.25 pc and $\sim$70 km~s$^{-1}$, respectively. Then we can derive a dynamic timescale of $t_{dyn}=\frac{L}{V_{char}}=8.5\times10^{3}$~yr, where $V_{char}=P/M$ is $\sim$29 km~s$^{-1}$, the averaged characteristic outflow velocity of the blue and red outflows. The outflow mass loss rate of MM6 is $\dot{M}_{out}=\frac{M_{out}}{t_{dyn}}=1.6\times10^{-4}$ M$_{\sun}$yr$^{-1}$. It is hard to estimate the dynamic timescale for the outflow of MM8/F due to its complicated structures. Here we simply assume its timescale of $1\times10^{4}$~yr, as also used in \cite{liu11}. The dynamic timescale of MM7 is assumed the same as MM8/F. The total outflow masses of MM7 and MM8/F are $\sim$1.2 and $\sim$14 M$_{\sun}$. Therefore, the outflow mass loss rates of MM7 and MM8/F are $1.2\times10^{-4}$ and $1.4\times10^{-3}$, respectively.

If outflows are powered by winds driven by accretion disks, the
outflow force ($F_{out}$) is related to the mass accretion rate
($\dot{M}_{acc}$) as given in the following equation obtained from the
principle of momentum conservation \citep{bon96,norb00,mck03,keto03,zhang01,zhang05}:

\begin{equation}\label{eq_finalmdot}
\dot{M}_{\rm acc} = \frac{1}{f_{\rm ent}} \, \frac{\dot{M}_{\rm acc}}{\dot{M}_w} \frac {1}{V_w} \, F_{flow}.
\end{equation}

We assume a typical jet/wind velocity of $V_w \sim 500$ km~s$^{-1}$ \citep{lam95}. $\dot{M}_w$ is the
wind/jet mass loss rate. Models of jet/wind formation predict, on average, $\dot{M}_w / \dot{M}_{acc}\sim$ 0.1 \citep{shu94,pell92,ward93,bon96}.  The entrainment efficiency is typically $f_{\rm ent} \sim 0.1-0.25$. Here we take 0.25. The outflow force $F_{flow}$ is calculated as:

\begin{equation}
F_{flow}=\frac{P}{t_{dyn}}
\end{equation}

Here we use the momentum P derived from SiO (5-4) considering abundance enhancement for MM6 and MM7/G. The momentum P of MM8/F is derived from CO (2-1) by applying optical depth correction. The total mass accretion rates for MM6, MM7/G, MM8/F are about 3.9$\times10^{-4}$, 5.7$\times10^{-5}$ and 2.2$\times10^{-3}$ M$_{\sun}$~yr$^{-1}$. The uncertainties of mass accretion rates are mainly caused by the uncertainties of outflow parameters as discussed above. Therefore, they should be treated with respectful caution. However, the mass accretion rates derived here are comparable to those in other outflow studies toward high-mass star forming regions \citep[e.g., ][]{zhang05,qiu09,liu11,liu16c}.

\section{Discussions}

\subsection{Sequential high-mass star formation in the G9.62 clump}

\subsubsection{High-mass protostellar cores}

Five of the twelve dense cores are obvious high-mass protostars: MM4/E, MM6, MM7/G, MM8/F and MM11/D. CH$_{3}$OH is steadily produced in ice mantles of dust grains until the richness of molecular composition reaches its maximum in the hot molecular core (HMC) phase and declines for the UC H{\sc ii} stage, because CH$_{3}$OH is destroyed by the UV radiation from the ionizing stars \citep{gern14}. Therefore, CH$_{3}$OH can be used as chemical diagnostics of the evolutionary stages of high-mass star forming clumps \citep{gern14}. CH$_{3}$OH v$_{t}$=1 ($6_{1,5}-7_{2,6}$) has an upper level energy of $\sim$374 K. In addition, molecular vibrational transitions are thought to be excited by radiation from high mass stellar objects. Therefore, the detection of CH$_{3}$OH v$_{t}$=1 ($6_{1,5}-7_{2,6}$) line should indicate that the YSO has significantly heated its surroundings and thus the YSO should be at hot core phase or UC H{\sc ii} regions. Outflows are powerful tools to date the accretion history of young protostellar objects \citep{arce07}. In addition, the outflow collimation and morphology also changes with time. The youngest outflows are highly collimated while older ones present much lower collimation factors \citep{arce07}. Therefore, outflows can also be used as a clock for core evolution. Below we classify the evolutionary stages of the five protostellar cores based on their CH$_{3}$OH v$_{t}$=1 ($6_{1,5}-7_{2,6}$) line and outflow properties.

MM4/E and MM11/D are the two most evolved cores because they show strong centimeter continuum emission \citep{gar93,tes00}, and do not drive molecular outflows, indicating that their accretion may have been halted by stellar feedback (e.g., radiation). MM11/G is an UC H{\sc ii} region excited by a B0.5 star (Hofner et al. 1996). MM4/E shows much stronger CH$_{3}$OH v$_{t}$=1 ($6_{1,5}-7_{2,6}$) emission than MM11/G. Since CH$_{3}$OH abundance is expected to reach its maximum in the hot molecular core (HMC) phase and decline for the UC H{\sc ii} stage \citep{gern14}, MM4/E may be still at an earlier phase (late hot core phase or hyper compact (HC) H{\sc ii} region) than the UC H{\sc ii} region MM11/G. MM7/G and MM8/F were also detected in centimeter continuum \citep{tes00} and show strong CH$_{3}$OH v$_{t}$=1 ($6_{1,5}-7_{2,6}$) emission. They are less evolved than MM4/E and MM11/D  because they are driving energetic outflows, suggesting that they are still at accretion phase.

Besides CH$_{3}$OH v$_{t}$=1 ($6_{1,5}-7_{2,6}$) line, we also detected CH$_{3}$OCHO v=1 ($17_{4,13}-16_{4,12}$) (rest frequency: 217.31263 GHz) emission toward MM4/E, MM8/F and MM11/D. Complex organic molecules are good indicator of hot cores \citep{kurt00,qin10}. The detection of CH$_{3}$OCHO v=1 ($17_{4,13}-16_{4,12}$) emission at MM4/E, MM8/F, MM11/D suggest that the three sources show hot core chemistry. CH$_{3}$OCHO v=1 ($17_{4,13}-16_{4,12}$) line was not detected toward MM7/G, indicating that MM7/G may be colder and less evolved than MM8/F. MM7/G may be at an early stage of hot core evolution. MM6 was not detected in centimeter continuum and also shows no CH$_{3}$OH v$_{t}$=1 ($6_{1,5}-7_{2,6}$) emission, suggesting that it is much younger than the other high-mass protostars in the G9.62 clump. MM6 seems to be at high-mass protostellar object (HMPO) phase because it is driving a very collimated outflow. In conclusion, we witness sequential high-mass star formation taking place within the G9.62 clump based on their evolutionary stages though we cannot accurately determine their ages. The evolutionary sequence of the 5 high-mass protostellar cores are MM6 (HMPO)$\rightarrow$MM7/G (early HMC)$\rightarrow$MM8/F (HMC)$\rightarrow$MM4/E (late HMC or HC H{\sc ii})$\rightarrow$MM11/D (UC H{\sc ii}).

\subsubsection{Candidates for high-mass starless cores}

The other cores (MM1, MM2, MM3, MM5, MM9, MM10 and MM12) were not detected in centimeter continuum and do not drive molecular outflows, indicating that they may be at earlier phases than the high-mass protostellar cores mentioned in section 4.1.1. MM1 shows broad SiO (5-4) emission but no CH$_{3}$OH v$_{t}$=1 ($6_{1,5}-7_{2,6}$) emission. MM1 has the largest radius among the 12 dense cores and lies in the interface between the H{\sc ii} region ``C" and the G9.62 clump. It may be just a remnant core in the envelope of the H{\sc ii} region ``C" \citep{liu11}. Its broad SiO emission is likely caused by the H{\sc ii} region shocks. No SiO (5-4) or CH$_{3}$OH v$_{t}$=1 ($6_{1,5}-7_{2,6}$) emission is detected toward MM2, and MM3, indicating that they are very likely starless cores. MM5 is located to the south of the UC H{\sc ii} region MM4/E but shows much narrower SiO (5-4) emission than MM4/E. Its CH$_{3}$OH v$_{t}$=1 ($6_{1,5}-7_{2,6}$) shows double-peaked profile with the blueshifted one stronger than the redshifted one, a typical ``blue profile", indicating that the core may be in collapse \citep{zhou93}. Its CH$_{3}$OH v$_{t}$=1 ($6_{1,5}-7_{2,6}$) emission may be caused by the external heating from MM4/E. MM9 is the densest core among the 12 cores. It shows no CH$_{3}$OH v$_{t}$=1 ($6_{1,5}-7_{2,6}$) emission. The SiO (5-4) emission toward MM9 exhibits line wings. However, since MM9 is at the vicinity of the energetic outflow lobes of MM8/F, its SiO (5-4) emission may be contaminated by the outflow emission of MM8/F. Considering its large density, MM9 seems to be a prestellar core on the verge of star formation. No CH$_{3}$OH v$_{t}$=1 ($6_{1,5}-7_{2,6}$) emission is detected toward MM10. However, MM10 shows very broad SiO (5-4) emission with blueshifted line wings because it is located at the center of the blueshifted outflow lobe (see section 3.4). Therefore, the broad SiO (5-4) emission at MM10 is not from its core emission. MM10 may be also a starless core. MM12 has no CH$_{3}$OH v$_{t}$=1 ($6_{1,5}-7_{2,6}$) emission. Its SiO (5-4) emission is only detected with S/N$\sim$4 $\sigma$ and has much narrower linewidth ($\sim$1.7 km~s$^{-1}$) than other cores. The seven cores (MM1, MM2, MM3, MM5, MM9, MM10 and MM12) are good candidates for starless cores, which need to be confirmed by more sensitive higher angular resolution observations.

In contrast to the ``Competitive Accretion" model, the ``Turbulent Core Accretion" model predicts the existence of massive starless cores. Finding massive starless cores is a very challenging problem. Recent high angular resolution interferometric observations have identified several candidates of massive starless cores \citep[e.g.,][]{cyga14,cyga17,kong17}. However, none of them have been fully confirmed \citep{kong17}. Among the seven starless core candidates in the G9.62 clump, five cores (MM1, MM3, MM5, MM9, MM10) have masses larger than 10 M$_{\sun}$ and volume densities larger than 10$^{6}$ cm$^{-3}$. Their masses may be underestimated due to the large dust temperature (35 K) used in mass calculation. In contrast, most starless cores in nearby clouds or IRDCs have temperature of 5-20 K \citep{and10,ward16,tan13,kong17}. If we adopt a dust temperature of 20 K, the masses of the five cores will become two times larger (see the values in parentheses in Table \ref{corepara}) and above 25 M$_{\sun}$. In this case, they will have ability to form high-mass stars (M$>$8 M$_{\sun}$) if assuming a local star formation efficiency at the core level of 30\% \citep{and14}. MM9 is particularly interesting due to its large volume density ($\sim10^{7}$ cm$^{-3}$). Its virial mass derived with the line width ($\sim$3 km~s$^{-1}$) of SiO (5-4) is $\sim$29 M$_{\sun}$, which is comparable to the core mass ($\sim$27 M$_{\sun}$) estimated with a dust temperature of 20 K, indicating that MM9 may be in virial equilibrium. The most massive starless core candidate in \cite{kong17} has a mass of $\sim$70 M$_{\sun}$ assuming a dust temperature of 10 K. Adopting a dust temperature of 10 K, the core mass of MM9 is also around 70 M$_{\sun}$, which is much larger than its virial mass, indicating that MM9 should be in collapse and may be on the verge of star formation.

We should stress that the properties of the starless core candidates in the G9.62 clump are not well known from present data. Future multiple wavelength high resolution and high sensitivity continuum and line observations with ALMA or NH$_{3}$ line observations with JVLA will be helpful to better constrain the dynamical and chemical properties of these cores.

\subsection{Fragmentation analysis}

In this section, we investigate the fragmentation of the G9.62 clump as well as the interaction between the evolved H{\sc ii} regions (``B" \& ``C") and  the G9.62 clump.

Molecular gas pressure inside the G9.62 clump can be expressed as:
\begin{equation}
\frac{P_{mol}}{k}=nT_{eff}
\end{equation}

Where n=$(9.1\pm0.7)\times10^{4}$ cm$^{-3}$ is the mean number density of the G9.62 clump, k is the Boltzmann constant, and $T_{eff}$ is effective kinetic temperature $T_{eff}=\frac{C_{s,eff}^{2} \mu m_{H}}{k}$, where $\mu$=2.37 is the mean molecular weight. Here $C_{s,eff}$ is an effective sound speed including
turbulent support,
\begin{equation}
C_{s,eff}=[(C_{s})^{2}+(\sigma_{NT})^{2}]^{1/2}
\end{equation}
where $C_{s}$ is the thermal sound speed and $\sigma_{NT}$ is the non-thermal one dimensional velocity dispersion. The thermal sound speed
is related with the kinetic temperature T$_{k}$ as following: $C_{s}=(kT_{k}/\mu m_{H})^{1/2}$. Assuming T$_{k}$ equaling to the dust temperature of 35 K, the sound speed is 0.35 km~s$^{-1}$. The non-thermal one dimensional velocity dispersion $\sigma_{NT}$ can be calculated as follows:
\begin{equation}
\sigma_{NT} =\sqrt{\sigma_{C^{18}O}^{2}-\frac{kT_{k}}{m_{C^{18}O}}}
\end{equation}
and
\begin{equation}
\sigma_{C^{18}O}=\frac{\Delta V}{\sqrt{8ln(2)}}
\end{equation}
with $m_{C^{18}O}$ being the mass of C$^{18}$O, $\Delta$V=3.5 km~s$^{-1}$ is the linewidth of C$^{18}$O (3-2). The derived $\sigma_{NT}$ and $C_{s,eff}$ are 1.48 and 1.52 km~s$^{-1}$, respectively. Therefore, the effective temperature $T_{eff}$ and molecular gas pressure $\frac{P_{mol}}{k}$ are $\sim$670 K and $(6.1\times0.5)\times$10$^{7}$ K~cm$^{-3}$.

The ionized gas pressure inside the radio source ``B" is the sum of thermal pressure and
turbulent pressure:
\begin{equation}
\frac{P_{i}}{k}=2n_{e}T_{e}+n_{e}\mu_g m_{H}\sigma_{NT}^2/k
\end{equation}
Here $\mu_g$=1.4 considering both hydrogen and helium gas. We adopt the electron density n$_{e}$ of 1.6$\times10^{3}$ cm$^{-3}$ and an electron temperature T$_{e}$ of 10000 K \citep{gar93}.
The $\sigma_{NT}$ can be derived with the [Ne {\sc ii}] line as for C$^{18}$O (3-2). The sound speed ($C_{i}$) and non-thermal velocity dispersion ($\sigma_{NT}$) in radio source ``B" are 9.1 and 5.3 km~s$^{-1}$. The ionized gas pressure is $\sim$4.0$\times$10$^{7}$ K~cm$^{-3}$, which is comparable to the molecular gas pressure inside the G9.62 clump, suggesting the pressure from the evolved H{\sc ii} regions may still help confine the G9.62 clump together with gravity.

In addition, the PDR of radio source ``B" traced by 8 $\micron$ emission (see Figure 1) shows very dense contours in the region adjacent to the G9.62 clump, strongly suggesting that radio source ``B" is interacting with the G9.62 clump. The PDRs at the interfaces between H{\sc ii} regions ``B/C" and the G9.62 clump may have a larger pressure than the molecular gas pressure because that the PDRs reside in the shock
induced by the ionizing front and has a much higher density. Therefore, the H{\sc ii} regions should be still compressing the G9.62 clump. As depicted by the yellow dashed curve in the left panel of Figure \ref{continuum}, the northern end of the G9.62 clump is very likely bent by the compression of the H{\sc ii} region ``C". The C$^{18}$O map in Figure \ref{C18O} also shows hint on interaction between the evolved H{\sc ii} regions and the G9.62 clump. Especially, as shown in Figure \ref{ds9}, widespread Spitzer/IRAC 4.5 $\micron$ emission is seen between the H{\sc ii} regions and the G9.62 clump, hinting at H{\sc ii} region shocks. G9.62+0.19 complex is a well-known cluster forming region, which is located in the 3 kpc arm (Scoville et al.1987; Holfner et al. 1994).  The radio sources in this region also have similar velocities and should be spatially connected. It is hard to imagine that they are projected on sky by chance. Therefore, the sequential high-mass star formation in the G9.62 clump is very likely caused by H{\sc ii} region shocks as also suggested by \cite{hof94}. However, the formation and fragmentation of the G9.62 clump may not be determined by the ``Collect and Collapse" process as discussed in Appendix C.

\begin{deluxetable}{cc}
\centering
\scriptsize
\tablecolumns{2} \tablewidth{0pc}
\tablecaption{Fragmentation in the G9.62 clump \label{fragpara}} \tablehead{ \colhead{M$_{core}$}  & \colhead{$\lambda_{core}$} \\
\colhead{(M$_{sun}$)} & \colhead{(pc)}}
\startdata
\multicolumn{2}{c}{Observed values} \\
\cline{1-2}
$\sim$20 & $\sim$0.08 \\
\cline{1-2}
\multicolumn{2}{c}{Thermal Jeans fragmentation\tablenotemark{a}} \\
\cline{1-2}
6.0$\pm$0.1 & 0.13$\pm$0.01 \\
\cline{1-2}
\multicolumn{2}{c}{Turbulent fragmentation\tablenotemark{b}} \\
\cline{1-2}
504$\pm$20 & 0.56$\pm$0.02 \\
\cline{1-2}
\multicolumn{2}{c}{Thermal Cylindrical fragmentation\tablenotemark{c}} \\
\cline{1-2}
6.6$\pm$0.5 & 0.10$\pm$0.01 \\
\cline{1-2}
\multicolumn{2}{c}{Turbulent Cylindrical fragmentation\tablenotemark{d}} \\
\cline{1-2}
430$\pm$30 & 0.40$\pm$0.03 \\
\enddata
\tablenotetext{a}{Only considering thermal Jeans instability by taking T=35 K, and n=$(9.1\pm0.7)\times10^4$ cm$^{-3}$, which were determined from SED fit in Appendix A. Jeans mass is $M_{\rm J} =
0.877 \, M_{\odot} \,
\left( \frac{T}{10\, \mathrm{K}} \right)^{3/2} \,
\left( \frac{n}{10^5 ~ \rm cm^{-3}} \right)^{-1/2}$;
Jeans length is $\lambda_{\rm J} =
0.066 ~ {\rm pc} \,
\left( \frac{T}{10\, \mathrm{K}} \right) \,
\left( \frac{n}{10^5 ~ \rm cm^{-3}} \right)^{-1/2}$ \citep{wang14}. }
\tablenotetext{b}{Considering turbulent Jeans instability by taking T=670 K, and n=$(9.1\pm0.7)\times10^4$ cm$^{-3}$.}
\tablenotetext{c}{Only considering thermal support in Cylindrical fragmentation with velocity dispersion $\sigma$=0.38 km~s$^{-1}$. Following \cite{wang14}, the central density of the cylinder on the verge of collapse is derived as
the average of the clump density ($(9.1\pm0.7)\times10^4$ cm$^{-3}$) and core density ($(4.4\pm0.6)\times10^6$ cm$^{-3}$), which amounts to $n_{c}\sim(2.3\pm0.3)\times10^{6}$ cm$^{-3}$. The fragment mass in cylindrical fragmentation is $M_\mathrm{cl} = 575.3 M_\odot
\left( \frac{\sigma}{1 ~ \rm km \, s^{-1}} \right)^3
\left( \frac{n_c}{10^5 ~ \rm cm^{-3}} \right)^{-1/2}$; The typical spacing in cylindrical fragmentation is $\lambda_\mathrm{cl} = 1.24\, {\rm pc}
\left( \frac{\sigma}{1 ~ \rm km \, s^{-1}} \right)
\left( \frac{n_c}{10^5 ~ \rm cm^{-3}} \right)^{-1/2}$ \citep{wang14}.}
\tablenotetext{d}{considering turbulent support in Cylindrical fragmentation with $\sigma$=1.53 km~s$^{-1}$ and $n_{c}\sim(2.3\pm0.3)\times10^{6}$ cm$^{-3}$.}
\end{deluxetable}

What determines the fragmentation of the G9.62 clump? Following \cite{wang12,wang14}, we derive the masses and separation of dense cores predicted in Jeans fragmentation and Cylindrical fragmentation. The results are presented in Table \ref{fragpara}. The uncertainties of the fragmentation analysis are mainly caused by density and temperature measurements. However, we noticed that the these uncertainties are negligible as shown in Table \ref{fragpara}.  The predicted dense core spacing ($\sim$0.4-0.6 pc) and mass ($\sim$400-500 M$_{\sun}$) in turbulent Jeans fragmentation or turbulent Cylindrical fragmentation are much larger than the observed mean dense core spacing ($\sim$0.1 pc) and mass ($\sim$20 M$_{\sun}$). However, in thermal Jeans fragmentation or thermal Cylindrical fragmentation, the predicted dense core spacing ($\sim$0.1 pc) is very consistent with the observed value ($\sim$0.08 pc). The predicted dense core mass ($\sim$6.0 M$_{\sun}$) is about one third of the mean dense core mass ($\sim$20 M$_{\sun}$). However, we noticed that the predicted core mass is comparable to the masses ($\sim$4-15 M$_{\sun}$) of the most quiescent cores (e.g., MM2, MM3, MM5, MM9 and MM12) in the G9.62 clump. The more evolved cores (e.g., MM4, MM6, MM8, MM11) in the G9.62 clump have masses ($\geq$30 M$_{\sun}$) much larger than the predicted dense core mass ($\sim$6.0 M$_{\sun}$) because they have accumulated significant amounts of additional mass in their accretion phase. Taking MM8/F for example, in 10$^{4}$ yrs, it could have accumulated additional mass of 22 M$_{\sun}$ if assuming a constant accretion rate of 2.2$\times10^{-3}$ M$_{\sun}$~yr$^{-1}$. Therefore, thermal Jeans instability can nicely explain the fragmentation in the G9.62 clump. In contrast to IRDC clumps \citep{zhang09,zhang11,wang11,wang12,wang14,zhang15}, turbulence support seems not to govern the fragmentation near the sonic scale ($\sim$0.1 pc) around cores in the G9.62 clump. It seems that the external heating from the evolved H{\sc ii} regions (``B" \& ``C") and internal heating (and/or outflows) from the YSOs (e.g., ``MM11/D", ``MM4/E", ``MM8/F") have greatly increased the thermal Jeans masses in the G9.62 clump and suppressed further fragmentation.

The increased thermal Jeans mass may lead to a lack of low-mass protostellar population. As mentioned in section 3.1, eleven cores have masses larger than 10 M$_{\sun}$ and only one has masses smaller than 10 M$_{\sun}$. Despite a five $\sigma$ mass sensitivity of 0.02 M$_{\sun}$ (assuming T$_{d}$=35 K and $\beta$=1.5), there is indeed a significant lack of a widespread low-mass protostellar population in the G9.62 clump. A lack of a distributed low-mass protostellar population was also witnessed in the clump G28.34+0.06 P1 \citep{zhang15}. They argued that low-mass stars form at a later stage after the birth of more massive protostars in a protocluster \citep{zhang15}. In contrast, the lack of low-mass protostellar population in the G9.62 clump is more likely due to the stellar feedback from the H{\sc ii} regions (``B" and ``C"), which suppresses further fragmentation and successive low-mass star formation.

\begin{figure*}[tbh!]
\centering
\includegraphics[angle=90,scale=0.6]{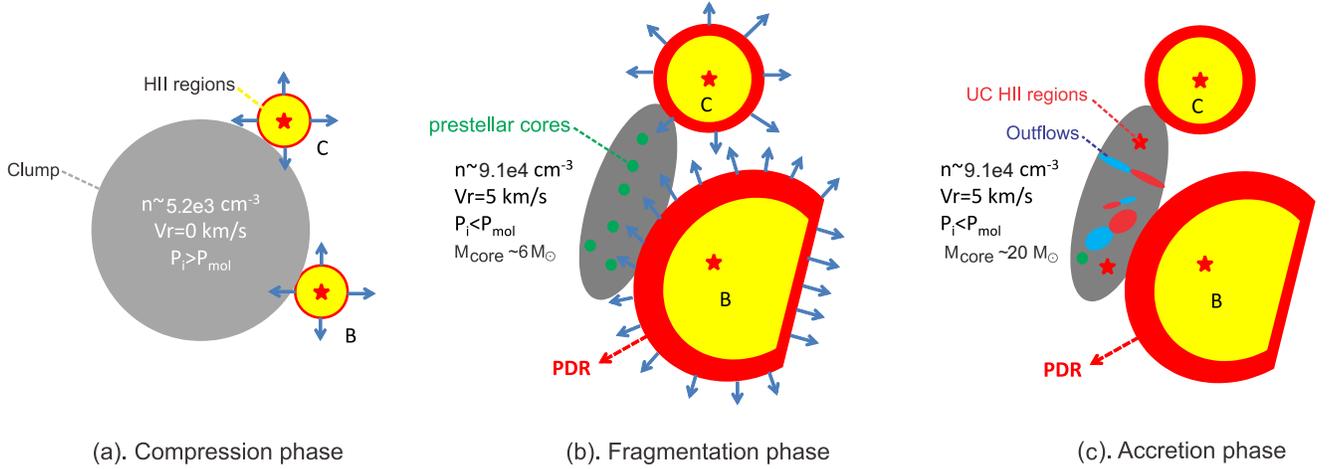}
\caption{A cartoon shows how the H{\sc ii} regions interact with a self-gravity clump (the G9.62 clump). Left panel: Compression phase. The unperturbed natal clump having a radius of $\sim$1.3 pc and a mean number density of 5.2$\times10^{3}$ cm$^{-3}$ would undergo globally collapse without external compression. The relative velocity (V$_{r}$) between the ionized gas and molecular gas at the initial stage is $\sim$0 km~s$^{-1}$. The ionized gas pressure ($P_{i}$) is larger than molecular gas pressure ($P_{mol}$). The shocks induced from nearby H{\sc ii} regions will compress and externally heat the clump from the west as they expand. Due to the external compression, the natal clump collapse to a filamentary (or Cylindrical) clump. The relative velocity (V$_{r}$) between the ionized gas and molecular gas becomes $\sim$5 km~s$^{-1}$ due to shocks. The ionized gas pressure ($P_{i}$) will decrease and the molecular gas pressure ($P_{mol}$) will increase in the compression phase. Middle panel: Fragmentation phase. The clump further fragments into dense cores with typical mass of 6 M$_{\sun}$ due to thermal instability. Right Panel: Accretion phase. The dense cores collapse to form stars and accrete more masses from the natal clump. The feedback (e.g., radiation, outflows) of the star formation further suppresses the fragmentation in the clump.\label{cart}  }
\end{figure*}

Although the formation and fragmentation of the G9.62 clump is not governed by the ``Collect and Collapse" process (see Appendix C), the expansion of the H{\sc ii} regions (``B" and ``C") may still have great effects on the G9.62 clump through external compression and heating. As shown in Figure \ref{cart}, here we propose a scenario to explain the interaction between the H{\sc ii} regions (``B" \& ``C") and the G9.62 clump:

(i). Compression phase.

The unperturbed self-gravity natal clump is in round shape and has a radius of $\sim$1.3 pc and a mean H$_{2}$ number density of 5.2$\times10^{3}$ cm$^{-3}$ (see Appendix C). At the initial stage, the ionizing pressure of the adjacent H{\sc ii} regions (``B" \& ``C") is much larger than the molecular gas pressure in the clump. The shocks induced by the ionizing front will externally compress and heat the clump from its western side, reshaping it to a dense filamentary structure.

(ii). Fragmentation phase.

The H{\sc ii} regions stop expansion when their ionizing gas pressure becomes in equilibrium with the molecular gas pressure in less than 1$\times10^{5}$ yrs. At that moment, the clump further fragments into regularly spaced dense cores with typical mass of 6.0 M$_{\sun}$ and spacing of $\sim$0.1 pc due to thermal instability.

(iii). Accretion phase.

The dense cores will accumulate more masses from the natal clump in their accretion phase. The feedback (e.g., radiation, outflows) of the star formation suppresses the further fragmentation in the clump. The H{\sc ii} region shocks may induce the collapse of dense cores and lead to the sequential star formation within the G9.62 clump. The cores (e.g., MM4/E, MM11/D) near the clump edges are closer to the H{\sc ii} regions (``B" or ``C") and are more easily to collapse due to the perturbation of the H{\sc ii} region shocks. Therefore, they form new stars first. In contrast, the dense cores (e.g., MM6, MM7/G, MM8/F) near the clump center are more shielded and less affected by the shocks and may collapse later to form stars. Particularly, core MM9, which is located to the south of MM8/F and MM11/D and is further away from the H{\sc ii} regions, is more likely still at the prestellar core phase. If there is no shock, the dense cores in the G9.62 clump may be at similar evolutionary stages \textbf{as the same in} the IRDC clump G28.34+0.06, where all the cores drive collimated outflows \citep{wang12,zhang15}. The other cores (MM2, MM3, MM5, MM10) are located close to vicinity of already formed luminous high-mass protostars (``D" and ``E") and may be induced secondary cores.

\subsection{Mass-velocity diagrams of outflows}

\begin{figure}[tbh!]
\centering
\includegraphics[angle=0,scale=0.5]{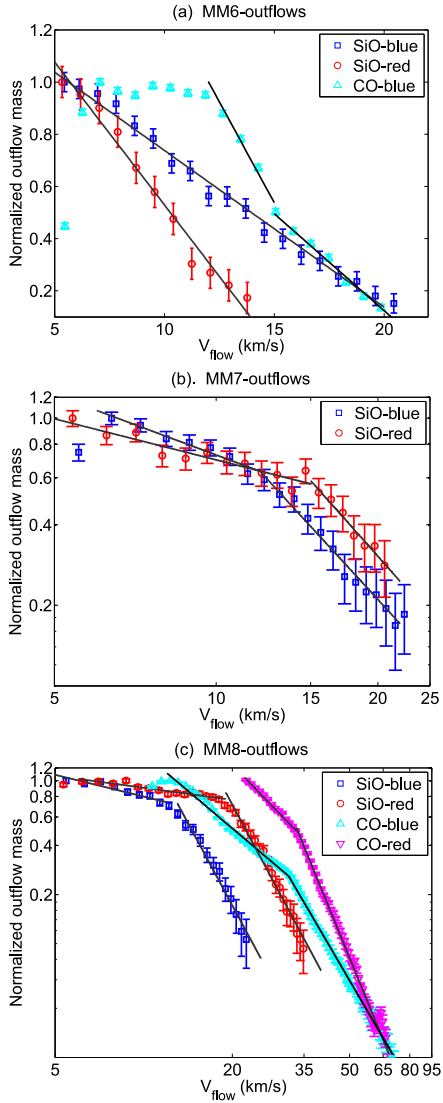}
\caption{Mass-velocity diagrams of outflows associated with MM6 (panel a), MM7/G (panel b) and MM8/F (panel c). The outflow masses derived from CO (2-1) or SiO (5-4) are normalized. The mass-velocity diagrams of outflows of MM6 were fitted with linear functions, while the outflows of MM7/G and MM8/F  were better fitted with broken power laws. The parameters of the fits are summarized in Table \ref{mvpara}.\label{MV}}
\end{figure}

\begin{deluxetable}{ccc}
\centering
\scriptsize
\tablecolumns{3} \tablewidth{0pc}
\tablecaption{The slopes of outflow mass-velocity relation \label{mvpara}} \tablehead{
\colhead{Flow Velocity intervals } & \colhead{a or $\gamma$} & \colhead{R$^{2}$}\\
\colhead{(km~s$^{-1}$)} }
\startdata
\multicolumn{3}{c}{Blueshifted SiO outflows of MM6\tablenotemark{a}}\\
\cline{1-3}
(5,14) & 0.11$\pm$0.01 & 0.98 \\
\cline{1-3}
\multicolumn{3}{c}{Redshifted SiO outflows of MM6\tablenotemark{a}}\\
\cline{1-3}
(5,25) & 0.06$\pm$0.01 & 0.99 \\
\cline{1-3}
\multicolumn{3}{c}{Blueshifted CO outflows of MM6\tablenotemark{a}}\\
\cline{1-3}
(12,15) & 0.14$\pm$0.01 & 0.97 \\
(15,21) & 0.07$\pm$0.01 & 0.99 \\
\cline{1-3}
\multicolumn{3}{c}{Blueshifted SiO outflows of MM7/G\tablenotemark{b}}\\
\cline{1-3}
(6,12) & 0.75$\pm$0.08 & 0.95 \\
(12,22) & 2.19$\pm$0.10 & 0.98 \\
\cline{1-3}
\multicolumn{3}{c}{Redshifted SiO outflows of MM7/G\tablenotemark{b}}\\
\cline{1-3}
(5,15) & 0.51$\pm$0.06 & 0.87 \\
(15,21) & 2.26$\pm$0.19 & 0.96 \\
\cline{1-3}
\multicolumn{3}{c}{Blueshifted SiO outflows of MM8/F\tablenotemark{b}}\\
\cline{1-3}
(5,12) & 0.44$\pm$0.05 & 0.91 \\
(12,22) & 3.36$\pm$0.17 & 0.97 \\
\cline{1-3}
\multicolumn{3}{c}{Redshifted SiO outflows of MM8/F\tablenotemark{b}}\\
\cline{1-3}
(6,18) & 0.25$\pm$0.02 & 0.91 \\
(18,35) & 3.38$\pm$0.13 & 0.98 \\
\cline{1-3}
\multicolumn{3}{c}{Blueshifted CO outflows of MM8/F\tablenotemark{b}}\\
\cline{1-3}
(12,31) & 1.53$\pm$0.03 & 0.99 \\
(31,70) & 3.10$\pm$0.03 & 1.00 \\
\cline{1-3}
\multicolumn{3}{c}{Redshifted CO outflows of MM8/F\tablenotemark{b}}\\
\cline{1-3}
(22,33) & 1.86$\pm$0.04 & 1.00 \\
(33,70) & 4.42$\pm$0.08 & 0.99 \\
\enddata
\tablenotetext{a}{Linear fit: M(v)$\propto$-a$\times$V}
\tablenotetext{b}{Power law fit: M(v)$\propto$V$^{-\gamma}$}
\end{deluxetable}

Outflow velocities decrease as the ambient cloud gas is swept up. A broken power law, $dM(v) / dv \propto v^{-\gamma}$ usually
exhibits in the mass-velocity diagrams of molecular outflows near young stellar objects
\citep{cha96,lad96,rid01,su04,arce07,qiu07,qiu09}, which may serve as a diagnostic
of the interaction of the outflow with the ambient gas. The slopes usually change at velocities between 6 and 12 km~s$^{-1}$ and become steeper at higher velocities \citep{arce07}. The breaks in the mass-velocity diagrams simply reflect a decrease of mass entrainment efficiency with increasing outflow velocities \citep{qiu09}. In recent MHD simulations, the breaks in the mass-velocity diagrams of protostellar outflows usually occur between 4 and $\sim$20 km~s$^{-1}$ \citep{li17}. In the simulation, protostars with a low
break velocity have either a weak outflow, generally due to their youth, or an outflow with a large inclination angle with respect to the line of sight \citep{li17}.

We calculate the outflow masses in each velocity channel and present the mass-velocity diagrams of the molecular outflows associated with MM6, MM7/G and MM8/F in Figure \ref{MV}. The slopes of the mass-velocity diagrams are summarized in Table \ref{mvpara}. Our findings are:

(i). The mass-velocity diagrams of outflows associated with MM6 significantly deviate from broken power laws. Instead, the mass-velocity diagrams of its SiO (5-4) outflows can be fitted with single linear function. The mass-velocity diagrams of its CO (2-1) outflows can be fitted with a broken linear function. Such mass-velocity diagrams may be due to the large inclination angle of outflows with respect to the line of sight. The outflows of MM6 seem to be parallel to the plane of sky and thus the mass-velocity diagrams are not accurately derived.

(ii). The mass-velocity diagrams of outflows associated with MM7/G and MM8/F can be well fitted with broken power laws. Interestingly, the breaks of blueshifted outflows usually occur at smaller velocities than redshifted outflows. The break velocities of the SiO blueshifted outflows associated with both MM7/G and MM8/F are around 12 km~s$^{-1}$. In contrast, the break velocities of the SiO redshifted outflows associated with MM7/G and MM8/F are around 15 and 18 km~s$^{-1}$, respectively. The break velocities of the CO blueshifted and redshifted outflows associated MM8/F are around 31 and 33 km~s$^{-1}$, respectively. The differences between blueshifted and redshifted outflows are not well known but most likes attributable to the projection effect, which prevents accurate determination of flow velocities possible.

(iii). The slope of the mass-velocity diagram may steepen with age and energy in the flow \citep{dow99,rich00,arce07}. The slope $\gamma$ of the high-velocity (V$_{flow}>$12 km~s$^{-1}$) SiO outflows associated with MM7/G is $\sim$2.2, which is smaller than the $\gamma$ ($\sim$3.4) of high-velocity SiO outflows associated with MM8/F, indicating that the outflows of MM7/G may be younger than the outflows of MM8/F \citep{dow99}.

(iv). The mass-velocity diagrams of SiO outflows associated with MM8/F break at much smaller velocities (12-18 km~s$^{-1}$) than CO outflows (31-33 km~s$^{-1}$). In addition, the mass-velocity diagrams of the outflows traced by SO ($8_{7}-7_{7}$), CS (7-6) and HCN (4-3) seem not to break at velocities up to 25 km~s$^{-1}$ \citep{liu11}. Why do SiO outflows break much earlier than other outflow tracers? It seems plausibly that SiO traces newly shocked gas while the other molecular lines (e.g., CO, SO, CS, HCN) mainly trace the ambient gas continuously entrained by outflow jets. The SiO outflow appears younger than the other molecular outflows (e.g., CO) and thus shows a lower
break velocity \citep{li17}. Different molecular tracers may be used to reveal the mass entrainment history of outflows, which need to be investigated thoroughly through detailed studies of outflow chemistry.

\subsection{Comparison with other high-mass star forming clumps}

Total infrared luminosities L$_{TIR}$ can well trace the star formation rates (SFRs; $\dot{M_{*}}$) in high-mass star forming clumps (HMSFCs). \cite{liu16b} found a tight linear relationship between L$_{TIR}$ and Clump masses M$_{clump}$ in a large sample of UC H{\sc ii} regions, suggesting a constant gas depletion time of $\sim$100 Myr. The G9.62 clump has a L$_{TIR}$ of $(1.7\pm0.1)\times10^{6}$ L$_{\sun}$, corresponding to a SFR ($\dot{M_{*}}$) of $(2.5\pm0.1)\times10^{-4}$ M$_{\sun}$~yr$^{-1}$ \citep{kenn12}. The gas depletion time ($\tau_{dep}$) is $\tau_{dep}=\frac{M_{clump}}{\dot{M_{*}}}=11\pm1$ Myr, where $M_{clump}=2800\pm200 M_{sun}$ is the clump mass. The gas depletion time of G9.62 clump is about ten times shorter than the average gas depletion time of Galactic HMSFCs. The star formation efficiency probed by L$_{TIR}$/M$_{clump}$ in the G9.62 clump is enhanced by a factor of 10 when compared with other HMSFCs in \cite{liu16b}. The enhanced star formation in the G9.62 clump is likely due to the external compression from H{\sc ii} regions, which may have very positive effect on new star formation because compression by the H{\sc ii} regions can quickly push more gas to the originally dense clumps and may even increase the density of the clumps further.

Similar to the G9.62 clump, globally collapsing clumps (e.g., SDC335, G10.6-0.4, AFGL 5142) are also highly fragmented \citep{pere13,liu13a,liu16c}. However, those globally collapsing clumps are more mass concentrated and the dense cores in globally collapsing clumps are not aligned in a chain with regularly spacing as in the G9.62 clump. The most massive cores form at the centers of globally collapsing clumps and are always surrounded by a swarm of low-mass cores \citep{pere13,liu13a,liu16c}. In contrast, the dense cores in the G9.62 clump have very comparable masses. Highly fragmented clumps with regularly spaced dense cores have also been detected in filamentary infrared dark clouds \citep{zhang09,zhang11,wang11,wang12,wang14,zhang15}. In contrast to the G9.62 clump, where the fragmentation seems to be governed by thermal instability, the fragmentation in those IRDC clumps are thought to be governed by turbulent Jeans instability \citep{wang12,wang14}. In addition, the dense cores in those more quiescent IRDC clumps are more likely at similar evolutionary stages \citep{wang12,wang14,zhang15}, which is very different from the G9.62 clump where sequential high-mass star formation is taking place. Those differences between the G9.62 clump and other clumps (either globally collapsing clumps or IRDC clumps) may also attribute to the external compression from the H{\sc ii} regions. Only through statistical studies toward more clumps similar to the G9.62 clump can we thoroughly understand the effect of stellar feedback on new generation of star formation.

\section{Summary}

We present ALMA and NASA/IRTF observations of the G9.62+0.19 complex. Together with archived data from Spitzer, Herschel and JCMT, we studied the properties of the dense clump (the G9.62 clump) in the G9.62+0.19 complex and its interaction with more evolved H{\sc ii} regions. The main findings are:

(1). The G9.62 clump is embraced by the PDRs from the western evolved H{\sc ii} regions (``B" and ``C"). Radio sources ``B" and ``C" were detected together with a more evolved HII region (``A") in 12.8 $\micron$ [Ne {\sc ii}] emission, as well. The ionized gas traced by 12.8 $\micron$ [Ne {\sc ii}] emission shows different velocities from the molecular gas. The velocity difference is $\sim$5 km~s$^{-1}$.

(2). The 1.3 mm continuum from ALMA observations resolves the G9.62 clump into 12 dense cores (MM 1-12). Only 4 of the 12 dense cores are previously known. The others are newly detected. The masses of the dense cores range from 4 to 87 M$_{\sun}$ with a median mass of $\sim$20 M$_{\sun}$ given a dust temperature of 35 K. The dense cores are at very different evolutionary stages, spanning from starless core phase to UC H{\sc ii} phase. Particularly, five cores (MM1, MM3, MM5, MM9, MM10) are massive starless core candidates with masses larger than 25 M$_{\sun}$ if assuming a dust temperature of $\leq$ 20 K. The ALMA observations reveal sequential high-mass star formation happening within the G9.62 clump.

(3). Three outflows are identified in SiO (5-4) or CO (2-1) line emission. MM6 is driving a very young ($\sim8.5\times10^{3}$ yr) and collimated outflow. The outflow associated with MM7/G may have a very small inclination angle with respect to the line of sight. The outflow from the hot core MM8/F is very energetic with a maximum flow velocity up to $\sim$100 km~s$^{-1}$ in CO (2-1) emission. The estimated mass accretion rates for MM6, MM7/G, MM8/F are 3.9$\times10^{-4}$, 5.7$\times10^{-5}$ and 2.2$\times10^{-3}$ M$_{\sun}$~yr$^{-1}$, respectively. The SiO abundance seems to be enhanced by a factor of 10-40 in outflow regions.

(4). The mass-velocity diagrams of outflows associated with MM7/G and MM8/F can be well fitted with broken power laws. In contrast, the mass-velocity diagrams of outflows associated with MM6 can be better fitted with linear functions due to its large inclination angle with respect to the line of sight. The mass-velocity diagram of SiO outflow associated with MM8/F breaks much earlier than other outflow tracers (e.g., CO, SO, CS, HCN), suggesting that SiO traces newly shocked gas while the other tracers (e.g., CO, SO, CS, HCN) mainly trace the ambient gas continuously entrained by outflows.

(5). The fragmentation in the G9.62 clump may be governed by thermal instability. In contrast to IRDC clumps, turbulence support seems not to govern the fragmentation near the sonic scale ($\sim$0.1 pc) around cores in the G9.62 clump. The shocks from the expanding H{\sc ii} regions (``B", ``C") may reshape the natal G9.62 clump into a filament, which will further fragment to form new stars. The H{\sc ii} region shocks may be also responsible for the sequential star formation inside the G9.62 clump.

(7). The G9.62 clump shows higher star formation efficiency and shorter gas depletion time than other Galactic massive clumps. There is a
lack of a widespread low-mass protostellar population in the G9.62 clump. Our findings suggest that the stellar feedback from H{\sc ii} regions may be able to enhance the star formation efficiency and suppress the low-mass star formation in adjacent pre-existing massive clumps.

\section*{Acknowledgment}
\begin{acknowledgements}

Tie Liu is supported by KASI fellowship and EACOA fellowship. Tie Liu thanks the support from Peking University, Chinese Academy of Sciences, South America Center for Astrophysics (CASSACA) and Universidad de Chile during his stay in Chile. S.-L. Qin is supported by  the Joint Research Fund
in Astronomy (U1631237) under cooperative
agreement between the National Natural Science Foundation of
China (NSFC) and Chinese Academy of Sciences (CAS), by
the Top Talents Program of Yunnan Province (2015HA030). Ke Wang is supported by grant WA3628-1/1 of the German Research Foundation (DFG) through the priority program 1573 (``Physics of the Interstellar Medium''). This paper makes use of the following ALMA data: ADS/JAO.ALMA\#2013.1.00957.S. ALMA is a partnership of ESO (representing its member states), NSF (USA) and NINS (Japan), together with NRC (Canada), NSC and ASIAA (Taiwan), and KASI (Republic of Korea), in cooperation with the Republic of Chile. The Joint ALMA Observatory is operated by ESO, AUI/NRAO and NAOJ.

\end{acknowledgements}

\clearpage

\section{APPENDIX A\\
Spectral energy distribution (SED) of the whole G9.62 clump}

\renewcommand{\thefigure}{A\arabic{figure}}

\setcounter{figure}{0}

\renewcommand{\theequation}{A\arabic{equation}}

\setcounter{equation}{0}

\renewcommand{\thetable}{A\arabic{table}}

\setcounter{table}{0}

\subsection{Archived data from Herschel, JCMT/SCUBA, APEX/LABOCA and SEST/SIMBA}

We obtained publicly available Herschel data from the Herschel archive. The 70 and 160 $\micron$ continuum data were obtained with the Photodetector Array Camera \& Spectrometer \citep[PACS,][]{pog10}. The 250, 350 and 500 $\micron$ continua were observed with the Spectral and Photometric Imaging Receiver \citep[SPIRE,][]{gri10}. The angular resolutions at 70, 160, 250, 350 and 500 $\micron$ wavelengths are about 6$\arcsec$, 12$\arcsec$, 18$\arcsec$, 25$\arcsec$, and 36$\arcsec$, respectively. The flux densities of the G9.62 clump at Herschel bands were determined by \cite{moli16}. The source extraction at each Herschel band was performed with the CuTEx algorithm with locally estimated background levels removed \citep{moli16}.The details of caveats are provided in \cite{moli16}. The 450 $\micron$ and 850 $\micron$ continuum data were obtained from the JCMT archive (Project ID: m98bu37). The JCMT/SCUBA observations were conducted in August 1998. The beam sizes of SCUBA at 450 $\micron$ and 850 $\micron$ are 9.8$\arcsec$ and 14.4$\arcsec$, respectively. The flux density of the G9.62 clump at each JCMT/SCUBA band was obtained from 2-D Gaussian fit. We also used APEX/LABOCA 875 $\micron$ continuum images \citep{csen14,csen16} and SEST/SIMBA 1.2 mm continuum image \citep{fau04}. The flux density at APEX/LABOCA 875 $\micron$ was also obtained from 2-D Gaussian fit. The flux density at SEST/SIMBA 1.2 mm was determined by \cite{fau04} from 2-D Gaussian fit. For flux densities obtained from ground-based bolometric observations (JCMT/SCUBA, APEX/LABOCA and SEST/SIMBA), we did not perform background substraction because those observations have already filtered out large scale extended emission in data reduction. All the flux densities are summarized in Table \ref{fluxpara}.

\begin{figure*}[tbh!]
\centering
\includegraphics[angle=-90,scale=0.8]{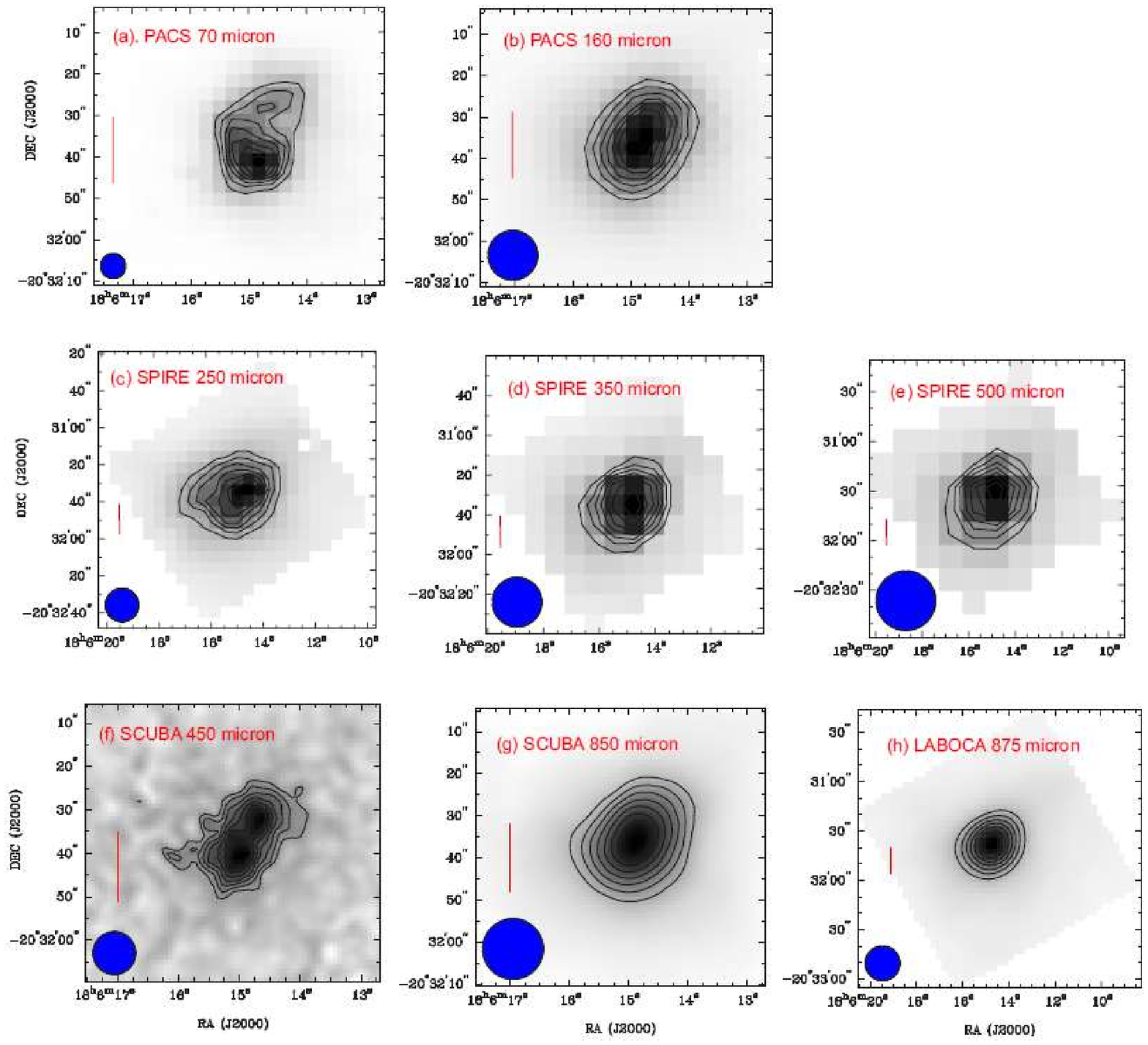}
\caption{Herschel and JCMT/SCUBA images of the G9.62 clump. The beams are shown in filled blue circles. The vertical red line represent a spatial scale of 0.4 pc. The contour levels are from 10\% to 90\% in steps of 10\% of the peak values. The peak values for (a) PACS 70 $\micron$, (b) PACS 160 $\micron$, (c) SPIRE 250 $\micron$, (d) SPIRE 350 $\micron$, (e) SPIRE 500$\micron$, (f) SCUBA 450 $\micron$, (g) SCUBA 850 micron, (h) LABOCA 875 $\micron$ are 4.9 Jy/pixel, 83.3 Jy/pixel, 44000 MJy/sr, 18000 MJy/sr, 4200 MJy/sr, 96 Jy/beam, 11 Jy/beam and 14 Jy/beam, respectively.\label{infra}  }
\end{figure*}

In Figure \ref{infra}, we present the continuum images of the G9.62 clump. The SEST/SIMBA 1.2 mm image is not shown here but included in \cite{fau04}. The relatively higher spatial resolution observations of Herschell/PACS 70 $\micron$ and JCMT/SCUBA 450 $\micron$ continuum maps resolve the G9.62 clump into two sub-clumps. While the 160, 250 (which is not shown here), 350, 500 and 850 $\micron$ continuum maps only reveal a single clump. The third column of Table \ref{fluxpara} presents the mean full width at half maximum (FWHM) deconvolved size (or effective radius). The radii of the two sub-clumps in 450 $\micron$ emission are $\sim$10$\arcsec$, or $\sim$0.25 pc at a distance of 5.2 kpc.

\subsection{SED fit}

\begin{deluxetable}{cccc}
\centering
\scriptsize
\tablecolumns{4} \tablewidth{0pc}
\tablecaption{The flux density of the G9.62 clump \label{fluxpara}} \tablehead{
\colhead{Wavelength} &  \colhead{F$_{int}$\tablenotemark{a}} &  \colhead{$\theta_{S}$\tablenotemark{a}} &  \colhead{Notes\tablenotemark{b}} \\
\colhead{$\micron$}  & \colhead{(Jy)} & \colhead{($\arcsec$)} }
\startdata
70 & 2512/1778 & 15.8/18.4 & 1\\
160 & 2886 & 22.7 & 1\\
250 & 1363 & 40.4 & 1\\
350 & 548.6 & 30 & 1\\
450 & 195/235 & 10.0/10.4 \\
500 & 211.6 & 42.4 & 1\\
850 & 32.5 & 21.4 \\
875 & 21.4 & 26 &2 \\
875 & 42.8 & 36.4 & 3\\
1200 & 11.4 & 29 &4 \\
\enddata
\tablenotetext{a}{The two sub-clumps in 70 $\micron$ and 450 $\micron$ images were fitted separately}
\tablenotetext{b}{1. Flux measurements at Herschel bands are from \cite{moli16}. 2. Flux measurement at 875 $\micron$ was performed only on APEX/LABOCA data \citep{csen14}. 3. Flux measurement at 875 $\micron$ was performed on APEX/LABOCA and Planck combined data \citep{csen16}. 4. Flux measurements at 1.2 mm are from SEST/SIMBA observations \cite{fau04}.}
\end{deluxetable}

\begin{figure}[tbh!]
\centering
\includegraphics[angle=-90,scale=0.3]{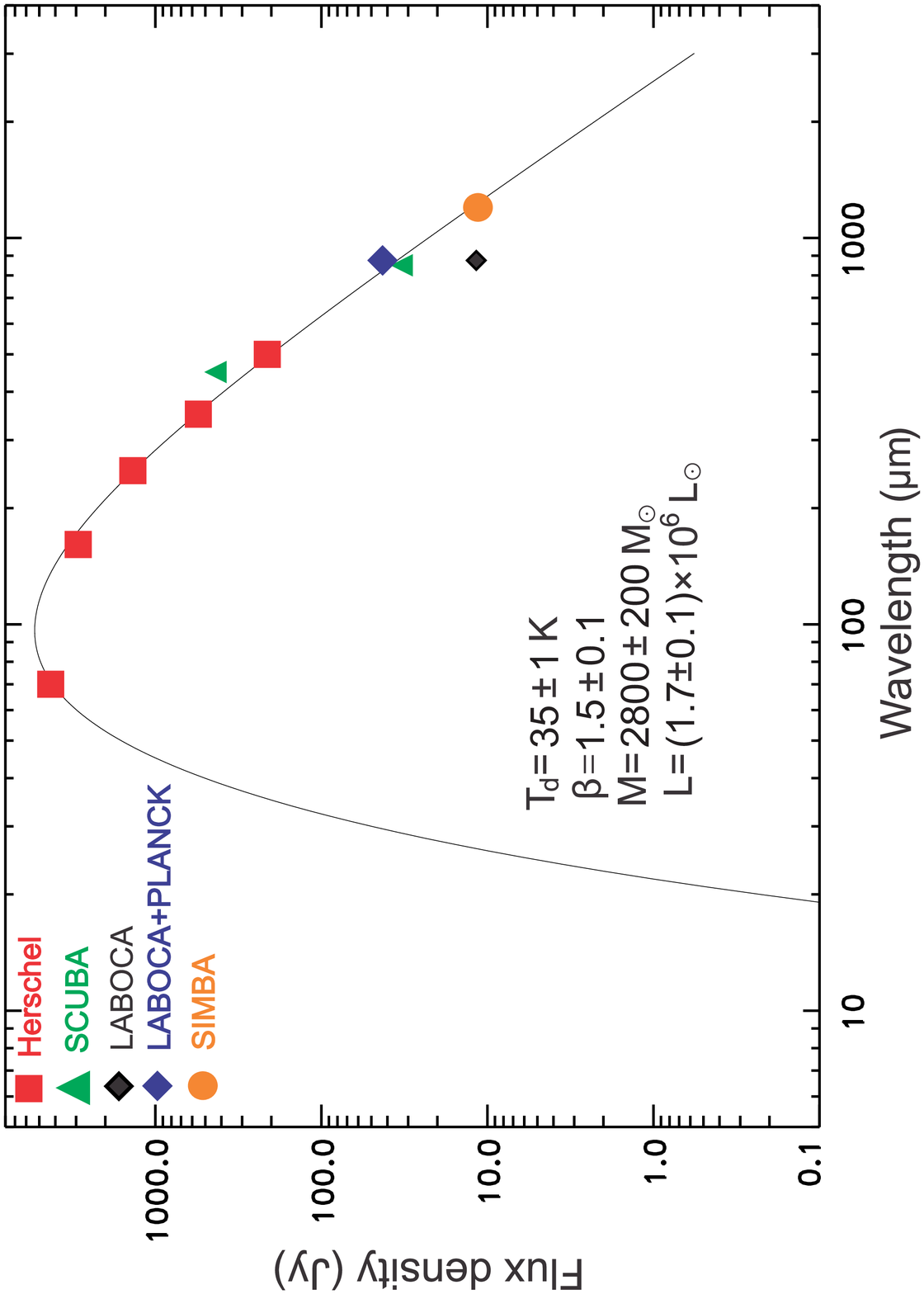}
\caption{The spectral energy distribution (SED) of the G9.62 clump. The best-fit SED is shown as a solid line. The flux at 70 and 450 $\micron$ are the sum of the two clumps. The errorbars of the data were set as 10\%, which is smaller than the mask sizes. \label{SED} }
\end{figure}

The spectral energy distribution (SED) of the G9.62 clump is shown in Figure \ref{SED}. The continuum
emission at the frequency $\nu$ from a thermal dust subtending a solid angle $\Omega$ with a dust temperature $T_{\rm d}$
and a total mass of gas and dust $M$, can be described as
\begin{equation}
\label{} S_{\nu} = B_{\nu}(T_{\rm
d})(1-e^{-\tau_{\nu}})\Omega
\end{equation}
where $S_{\nu}$ is the total flux density of the dust emission, $\tau_{\nu}$ is optical depth, and $B_{\nu}(T_d)$ is the Planck
function.

\begin{equation}
\tau_{\nu} = \kappa_{\nu}M/D^2\Omega = \kappa_{\nu}N_{H_{2}}\mu m_{H}
\end{equation}
where $\kappa_{\nu}$, D, $\mu$=2.37, $m_{H}$, and $N_{H_{2}}$ are the dust opacity per unit gas mass, distance, mean molecular weight, atomic hydrogen mass, and H$_{2}$ column density, respectively.
Assuming a gas to dust ratio of 100, the dust opacity per unit gas mass is $\kappa_{\nu}=\kappa_{0}(\frac{\nu}{\nu_{0}})^{\beta}$, where $\kappa_{0}=0.01$ cm$^{2}$g$^{-1}$ is the dust opacity at
$\nu_{0}$=230 GHz derived from \cite{oss94}.

We used a nonlinear least-squares method (the Levenberg--Marquardt algorithm coded within IDL) to fit the observed SED with equation (1) assuming a single dust component \citep{xue08}. Considering the uncertainties from calibration in all bands and background substraction in Hershel bands, we used 10\% uncertainties for flux densities in SED fit. The change (from 5\% to 20\%) of uncertainties does not affect the SED fit too much. The best fit model yielded a total dust and gas mass of 2800$\pm$200 M$_{\sun}$, a total luminosity of $(1.7\pm0.1)\times10^{6}$ L$_{\sun}$, a dust temperature of $35\pm1$ K and a dust opacity index ($\beta$) of 1.5$\pm$0.1. The low $\beta$ is consistent with previous estimations \citep{su05}. Considering that UC H{\sc ii} regions have formed in the G9.62 clump, another warm temperature component should contribute at shorter wavelengths. However, we failed to fit the SED with two temperature components due to the lack of data at near- and mid-infrared wavelengths. The G9.62 clump is not visible at shorter ($<$24 $\micron$) wavelengths due to the large extinction therein. However, as shown in the Figure \ref{SED}, the SED can be well fitted with one temperature component.

The G9.62 clump in SCUBA 850 $\micron$ and PACS 160 $\micron$ continuum images has a mean full width at half maximum (FWHM) deconvolved size (or effective radius) of $\sim$0.5 pc. Since SCUBA 850 $\micron$ and PACS 160 $\micron$ continuum images have better resolution than SPIRE and LABOCA images, and thus can better resolve the G9.62 clump, we take R=0.5 pc as the mean radius of the G9.62 clump. The mean H$_{2}$ volume density n can be calculated as $n=\frac{M}{\frac{4}{3}\pi R^{3}\mu m_{H}}$ and the mean column density N is N=$n\times2R$. The mean volume density and column density are $\sim(9.1\pm0.7)\times10^{4}$ cm$^{-3}$ and $\sim(2.8\pm0.2)\times10^{23}$ cm$^{-2}$, respectively.

\section{APPENDIX B\\
}

\begin{figure}[tbh!]
\centering
\includegraphics[angle=0,scale=0.4]{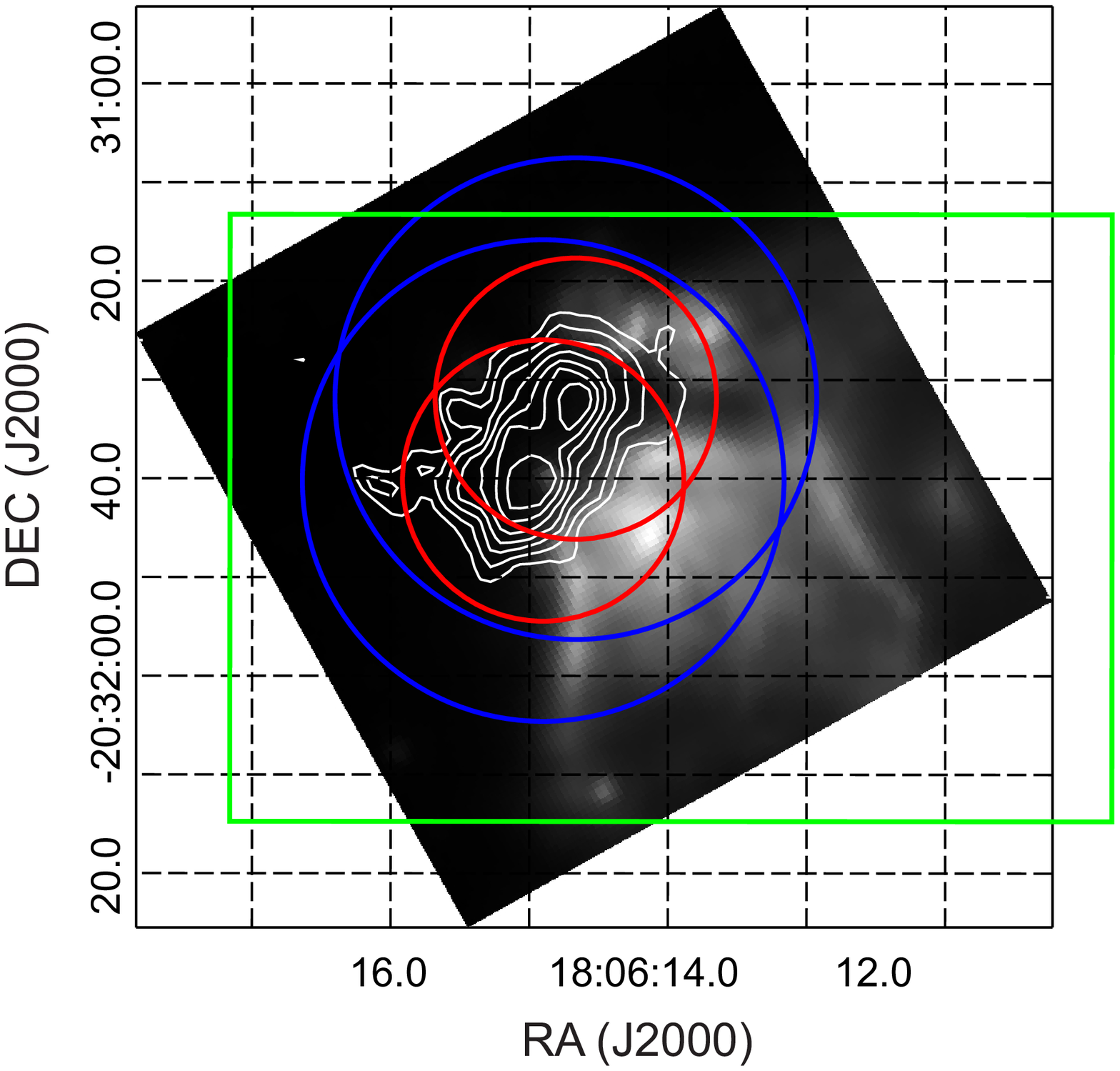}
\caption{ The Spitzer/IRAC 8 $\micron$ emission is also shown as gray-scale image. The white contours show the 450 $\micron$ emission. The contour levels are the same as in Figure \ref{ds9}. The green box marks the region covered by [Ne {\sc ii}] slides. The circles present the ALMA imaging area. The primary beams of ALMA 12-m array are shown in red circles. The primary beams of ALMA 7-m array are shown in blue circles. \label{obs}}
\end{figure}

Regions covered by ALMA and 12.8 $\micron$ [Ne {\sc ii}] observations are shown in Figure \ref{obs}.

\section{APPENDIX C\\
The ``Collect and Collapse" process
}

Is the formation and fragmentation of the G9.62 clump due to the ``Collect and Collapse" process as H{\sc ii} regions expand into homogenous interstellar medium \citep{el77,whit94a,whit94b}? \cite{whit94a,whit94b} found that the shocked shells swept up by H{\sc ii} regions, which expand into homogenous interstellar medium, may fragment to form new generation of stars within Myrs. We investigate the fragmentation process due to the ``Collect and Collapse" process assuming that the G9.62 clump was collected by the shock fronts induced from H{\sc ii} regions ``B" and ``C" in a homogeneous natal cloud following \cite{whit94a,whit94b}. The analysis of the ``Collect and Collapse" process requires initial hydrogen nuclei number density of $n_{0}$ and the initial sound speed of the natal cloud as well as total Lyman photon emission rate of the expanding H{\sc ii} regions \citep{whit94a,whit94b}. As outlined by the blue dashed circle in Figure \ref{C18O}, the unperturbed natal cloud may have a radius of $\sim$1.3 pc (or 50$\arcsec$). To obtain an approximate initial hydrogen nuclei number density of $n_{0}$ as input for the ``Collect and Collapse" model, we assume that the total mass of the G9.62 clump is distributed homogenously in a spherical unperturbed natal cloud (with radius of 1.3 pc), which give an estimated $n_{0}$ of $\sim$1.0$\times10^{4}$ cm$^{-3}$ (or mean molecular hydrogen number density of $\sim$5.2$\times10^{3}$ cm$^{-3}$). The total Lyman photon emission rate $\dot{N_{Lyc}}$ of ``B" and ``C" is $\sim$1.7$\times10^{48}$ s$^{-1}$ as estimated from radio continuum emission \citep{gar93}. We assume that the initial sound speed of unperturbed natal cloud is $\sim$0.26 km~s$^{-1}$ (T=20 K). Following \cite{whit94a,whit94b},  we obtain that the time (t$_{frag}$), the radius (R$_{frag}$) and column density through the shell (N$_{frag}$), the mean mass (M$_{frag}$) and initial separation of the resulting fragments (2r$_{frag}$), at which fragmentation of the shocked shell starts are 0.76 Myr, 1.55 pc, 1.6$\times10^{22}$ cm$^{-2}$, 24.6 M$_{\sun}$ and 0.52 pc, respectively. However, the R$_{frag}$ is larger than the radius of the H{\sc ii} region ``B" (15$\arcsec$, or $\sim$0.38 pc). The t$_{frag}$ is much longer than the dynamical age (t$_{dyn}$) of the H{\sc ii} region ``B" $t_{dyn}\sim\frac{R}{V_{S}}<\frac{0.38~pc}{10~km~s^{-1}}\sim3.7\times10^{4}$ yr. The mass of the resulting fragments (or dense cores) is roughly consistent with the masses of those dense cores in the ALMA observations. However, the separation (2r$_{frag}\sim$0.52 pc) is about five times larger than the averaged separation ($\sim$0.1 pc) of the ALMA dense cores. These inconsistences indicate that the formation and fragmentation of the G9.62 clump was not determined by the ``Collect and Collapse" process.

\end{document}